\documentclass[journal]{IEEEtran}

\usepackage{amsmath}
\usepackage{cite}
\usepackage{mathtools}

\usepackage{multicol}
\usepackage{soul}
\ifCLASSINFOpdf
\usepackage{graphicx}
\usepackage{multirow}
\usepackage{placeins}
\usepackage{txfonts}
\else
\usepackage{graphicx}
\fi
\usepackage{float}
\usepackage{color}
\ifCLASSOPTIONcompsoc
\usepackage[caption=false,font=footnotesize,labelfont=sf,textfont=sf]{subfig}
\else
\usepackage[caption=false,font=footnotesize]{subfig}
\fi
\usepackage{dblfloatfix}
\usepackage{algorithm}
\usepackage{algpseudocode}
\usepackage{breqn}
\usepackage{xcolor}

\begin{document}
\title{Novel Deep Learning Framework for Wideband Spectrum Characterization at Sub-Nyquist Rate}
		
 		\author{Shivam Chandhok, Himani~Joshi,  A V Subramanyam and Sumit J. Darak
	\thanks{Shivam Chandhok and Himani Joshi are joint first authors.}
					\thanks{This work is supported by the funding received from CSIR, India under SRF Scheme to Himani Joshi as well as DST INSPIRE and core research grant (CRG) awarded to Dr. Sumit J. Darak from DST-SERB, GoI.}
			\thanks{Shivam Chandhok is with Computer Science Department, 
				IIT-Hyderabad, India-502285 (e-mail: \{chandhokshivam\}@iith.ac.in) , Himani Joshi, Sumit J Darak and A V Subramanyam are with Electronics and Communications Department, 
				IIIT-Delhi, India-110020 (e-mail: \{himanij,sumit,subramanyam\}@iiitd.ac.in)}
}
	
	\maketitle
	\begin{abstract}
	
	Introduction of spectrum-sharing in 5G and subsequent generation networks demand base-station(s) with the capability to characterize the wideband spectrum spanned over licensed, shared and unlicensed non-contiguous frequency bands. Spectrum characterization involves the identification of vacant bands along with center frequency and parameters (energy, modulation, etc.) of occupied bands. Such characterization at Nyquist sampling is area and power-hungry due to the need for high-speed digitization. Though sub-Nyquist sampling (SNS) offers an excellent alternative when the spectrum is sparse, it suffers from poor performance at low signal to noise ratio (SNR) and demands careful design and integration of digital reconstruction, tunable channelizer and characterization algorithms. In this paper, we propose a novel deep-learning framework via a single unified pipeline to accomplish two tasks: 1)~Reconstruct the signal directly from sub-Nyquist samples, and 2)~Wideband spectrum characterization. The proposed approach eliminates the need for complex signal conditioning between reconstruction and characterization and does not need complex tunable channelizers. We extensively compare the performance of our framework for a wide range of modulation schemes, SNR and channel conditions. We show that the proposed framework outperforms existing SNS based approaches and characterization performance approaches to Nyquist sampling-based framework with an increase in SNR.  Easy to design and integrate along with a single unified deep learning framework make the proposed architecture a good candidate for reconfigurable platforms. 

	\end{abstract}
	
	\begin{IEEEkeywords}
 Deep learning, modulation classification, spectrum reconstruction, wideband spectrum digitization.
\end{IEEEkeywords}

		\section{Introduction}
	\label{intro}
	Automatic spectrum characterization (ASC) aims to blindly estimate the characteristics of the spectrum such as location of vacant and occupied bands, interference identification, modulation scheme along with direction-of-arrival of each band. The ASC has been discussed widely for military and cognitive radio (CR) applications. For military applications, it is used for electronic warfare, surveillance, signal jamming and target acquisition \cite{intro1}. Whereas in the CR applications, it is required for spectrum sharing \cite{nr} along with applications such as channel dependent rate adaptation \cite{intro2} and licensed user characterization to avoid the emulation attacks \cite{intro3}. Introduction of spectrum-sharing in 5G, radar, internet-of-things and next-generation networks demand base-station(s) with the ASC of wideband spectrum (ASCW) spanned over licensed, shared and unlicensed non-continuous frequency bands. 

	ASCW demands ultra high-speed Nyquist rate analog-to-digital converters (ADCs), computationally intensive filter-bank based channelizers to down-sample the desired frequency bands and characterization algorithms. To reduce the area, power and latency of ASCW, various sub-Nyquist sampling (SNS) methods have been explored which exploit the sparsity of a wideband spectrum to perform digitization via low-rate ADCs \cite{mcs,mwc,wcl,tsp}. However, they suffer from poor performance at low signal to noise ratio (SNR) and demands careful design and integration of reconstruction, channelizer and characterization algorithms.
	
	
	
	In the literature, ASC for narrowband spectrum has been well-explored \cite{survey} and various methods ranging from likelihood ratio classifier \cite{lr_amc}, feature-based classifier\cite{fb1,fb2} cyclostationary feature classifier \cite{cyclo_amc} 
    and machine learning classifiers such as k-nearest neighbor (KNN) \cite{intro1}, support vector machine (SVM) \cite{survey}, random forest (RF) \cite{rf} to the recent neural networks \cite{nn1} and deep learning (DL) \cite{ref1,ref2,ref3,ref4,ref5,ref6,tvt1,tvt2,tvt3} have been discussed. However, the performance analysis of these methods for ASCW has not been done yet. For instance, \cite{comsnet,sns_mc1,sns_mc2,sns_mc3,cssp} are the only works which consider ASCW, but they use random sampling based SNS, which is not feasible in hardware and state-of-the-art SNS approaches such as \cite{mcs,mwc} needs to be considered. 
    In addition, \cite{comsnet,sns_mc1,sns_mc2,sns_mc3,cssp} are based on individual frequency band characterization approach which demands complex signal conditioning,  tunable channelizer and multiplexer between reconstruction and characterization stages for multi-band characterization. Such blocks result in a heterogeneous architecture which is tedious to implement and integrate. From reconfigurability perspective, it offers limited flexibility due to fewer common functions between blocks. 
    
    
	
	
	In this paper, we propose a novel DL framework via a single unified pipeline to accomplish two tasks: 1)~Reconstruct the wideband spectrum signal sampled using state-of-the-art SNS methods, and 2)~Wideband spectrum characterization. Our main contributions are summarised below:

	\begin{figure*}[!t]
			\vspace{-0.2cm}
			\centering
			{\includegraphics[scale=0.4]{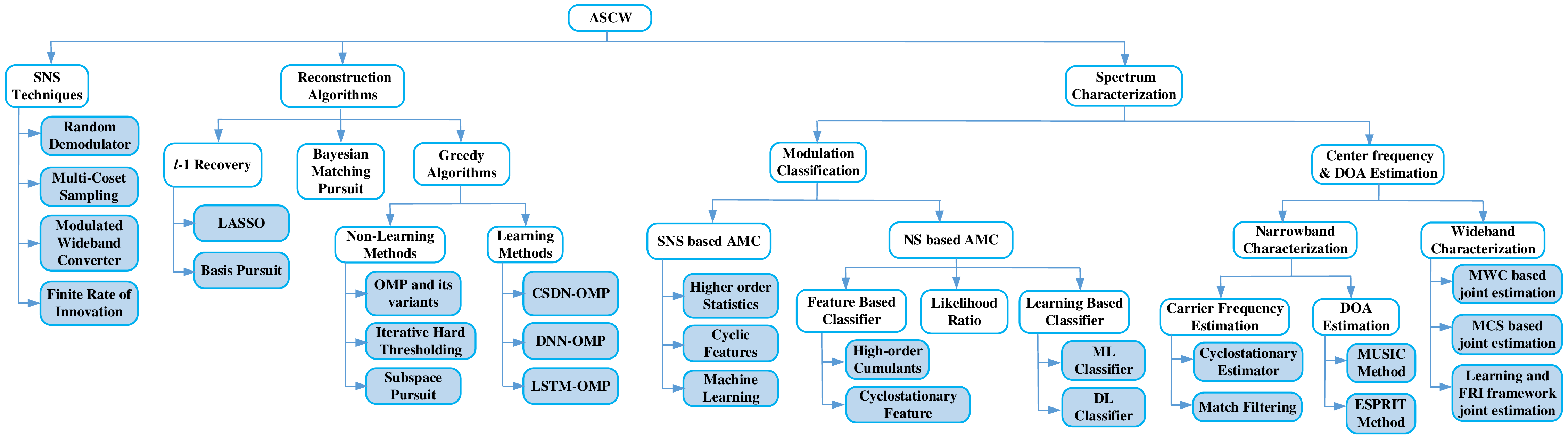}}						\vspace{-0.5cm}
			\caption{Various SNS, reconstruction and characterization approaches for automatic spectrum characterization of wideband spectrum (ASCW).}
			\vspace{-0.4cm}
			\label{summary}
		\end{figure*}
		

	\begin{enumerate}
	\item We offer a single unified pipeline of DL framework for ASCW, which performs direct characterization (band status and modulation scheme identification) on the multiband sub-Nyquist sampled wideband signal.
		
	\item  We replace a conventional digital reconstruction with new DL based approach. Considering the requirement of non-contiguous and reconfigurable SNS where parameters such as the number of bands, sparsity and sampling rate vary with time, the proposed architecture is agnostic to these parameters. 
	
	\item In terms of reconstruction performance, we show that the proposed non-iterative reconstruction approach offers better performance than conventional iterative orthogonal matching pursuit (OMP) approach \cite{omp} even though OMP needs sparsity information. 	
	
	\item We propose DL based modulation classifier that classifies all signals in the wideband spectrum simultaneously as compared to sequential single band classification in existing approaches \cite{dl_omp1,dl_omp2,dl_omp3}. Furthermore, the proposed approach is more sophisticated than a Velcro approach of stacking multiple classifiers in parallel, and we explore a new formulation of cross-entropy loss function to simultaneously classify multiple bands in wideband signal.
	

	

\item We develop and share new datasets which are the only publicly available datasets for SNS based ASCW \cite{google_drive_link}. We extensively compare the performance of our framework  for a wide range of modulation schemes, SNR and channel conditions. We show that the proposed framework outperforms existing SNS based approaches and characterization performance approaches to Nyquist sampling based framework with an increase in SNR.



	
	\end{enumerate}
	
	Easy to design and integrate due to fewer blocks  along  with a single unified DL  framework  to make it a  good  candidate  for  reconfigurable  platforms such as Zynq system-on-chip is one of the important aspects of the proposed work. The rest of the paper is organized as follows. The literature review of the existing ASCW methods is discussed in Section~II. Section~III describes the proposed end-to-end pipeline for ASCW. Section~IV discusses the proposed deep learning based wideband spectrum sensing (DLWSS) followed by the explanation of DL based modulation classification (DLMC) in Section~V. The Simulation set-up describing the dataset and tool used for DL based ASCW is described in Section~VI. Performance comparison of the proposed DLWSS and DLMC for ASCW is done in Section~VII followed by Conclusions in Section~VIII.
	
	\vspace{-0.2cm}
	\section{Literature Review}
	In this section, we review various state-of-the-art approaches involved in ASCW such as digitization, reconstruction and spectrum characterization. Please refer to Fig.~\ref{summary} for summary for various approaches in each domain.
	
	\label{Lit_Review}
	
		\vspace{-0.1cm}
	\subsection{Sub-Nyquist Sampling (SNS) Approaches}
	The Nyquist sampling based ASCW demands ultra-high speed ADCs, which are expensive and power-hungry. Various spectrum occupancy measurements indicated that the spectrum is sparse, and hence, it can be digitized at a fraction of the Nyquist rate. This led to various SNS approaches using low-speed ADCs such as multi-coset sampling (MCS) \cite{mcs}, modulated wideband converter (MWC) \cite{mwc} and finite rate of innovation (FRI) \cite{wcl,tsp}.

In MCS \cite{mcs}, each of the $K$ low-speed ADCs digitizes a wideband signal with a unique time-offset, and such offset is exploited to reconstruct the wideband signal from SNS samples. However, for ultra-wide spectrum, the offset is of the order of pico-seconds which is difficult to maintain in analog front-end (AFE). Furthermore, analog bandwidth of ADC is still high. The MWC overcomes these limitations with slightly higher complexity in AFE. In MWC, every ADC branch has a unique analog mixer followed by low pass filter and ADC.  The unique mixing function aliases all frequency bands at the baseband. Since the bandwidth of the aliased signal is same as the frequency band bandwidth, $B << f_{nyq}$, the analog bandwidth of the ADC is significantly reduced, i.e. $B~Hz$. Though conventional MCS and MWC can digitize only contiguous wideband spectrum, FRI based SNS \cite{wcl,tsp}, an extension of MWC, overcomes this limitation. As expected, FRI demands additional intelligence to chose which frequency bands to digitize and needs reconfigurable mixer in AFE. The design and configuration of the proposed ASCW approach are independent of SNS approach compared to \cite{sns_mc1,sns_mc2,sns_mc3}, which used a trivial random sampling approach.

		\vspace{-0.1cm}
	\subsection{Digital Reconstruction Approaches}
	After SNS, the next step is to reconstruct the wideband spectrum in the digital domain. Various reconstruction approaches have been explored in the literature and please refer to \cite{survey_rec} for more details. We can broadly classify these approaches into three categories: 1) Greedy algorithms \cite{greedy}, 2)~$l_1-$minimization algorithms \cite{l1_1,l1_2} and 3) Bayesian algorithms \cite{bp}. Among them, the greedy algorithm such as orthogonal matching pursuit, OMP \cite{omp} is widely used due to lower computational complexity. To improve the reconstruction performance, \cite{dl_omp1,dl_omp2,dl_omp3} replace the correlation step of OMP with DL methods i.e. convolutional deep stacking network (CSDN), multi-layer deep neural network (DNN) and long short term memory (LSTM), respectively. However, these methods follow the iterative approach and need prior knowledge of the number of busy bands. We aim to overcome these limitations in this paper.

			\vspace{-0.1cm}
	\subsection{Narrow Band Characterization}
	As discussed in Section I, the majority of the existing works focus on the narrowband characterization in which complex digital front-end (DFE) involving tunable channelizers are deployed to bring the desired frequency band to the baseband. {\color{black}For characterizing the carrier frequency, cyclostationary estimator and match filtering have widely studied \cite{survey_ss}. Furthermore, for direction of arrival (DOA) estimation, various variants of multiple signal classification (MUSIC) and estimation of signal parameters via rotational invariant technique (ESPRIT) algorithms are developed \cite{doa}. Whereas for determining the modulation schemes, various automatic modulation classification (AMC) techniques} such as likelihood ratio (LR) based classifiers \cite{lr_amc}, feature-based (FB) classifier \cite{fb1,fb2,cyclo_amc} and intelligent learning (IL) classifiers \cite{rf,nn1,ref1,ref2,ref3,ref4} are studied. LR based classifiers \cite{lr_amc} treat ASC as a multiple-composite hypothesis testing problem and parameters are determined by applying the maximum likelihood estimation (MLE) criteria. The drawback of this approach is that the accuracy depends on the knowledge of channel and noise model, which varies dynamically in the real environment. To overcome this, FB classification methods \cite{fb1,fb2} are studied. They analyze a variety of statistical features of the received spectrum such as first and second-order moments, cumulants\cite{fb1} and cyclostationary features\cite{cyclo_amc}. An extensive research has been focused on the usefulness and performance analysis of these methods \cite{lr_amc,fb1,fb2,cyclo_amc,rf,nn1,ref1,ref2,ref3,ref4}. 
	
	Recently, IL based AMC exploiting features along with learning algorithms such as SVM \cite{survey}, KNN \cite{intro1}, RF \cite{rf} and neural networks (NNs) \cite{nn1} have shown to offer a significant improvement in performance with limited prior knowledge of spectrum. Among them, NNs based ASC are state-of-the-art methods \cite{ref4}. NNs are DL models which work as function approximators and extract desirable patterns from the underlying relationships in data. In \cite{ref4}, it is shown that a simple convolutional NN (CNN) model with only two layers significantly outperforms expert based FB AMC methods. In \cite{ref1,ref2} authors adopt the principle architectures used for the task of image recognition for the ASC along with an extensive analysis to study the effect of network depth, filter sizes and the number of filters on the accuracy of classification. The representation of input signal for DL algorithms is analysed in \cite{ref3} and it has shown that the use of amplitude-phase samples with LSTM learning algorithm outperforms the real and imaginary sample based CNN ASC \cite{ref4}. Thus, the selection of an appropriate input format is important for ASC. We would also like to highlight the RadioML dataset \cite{radioML} is used widely for DL based ASC \cite{ref3,ref4}. However, this dataset along with existing state-of-the-art approaches \cite{ref3,ref4} are suitable only for narrowband ASC with Nyquist rate based digitization. {\color{black}Since there is no comprehensive datasets for wideband characterization using sub-Nyquist samples, we aim to generate such dataset for ASCW.}
	
			\vspace{-0.1cm}
	\subsection{Wideband Characterization}
	The extension of narrowband ASC to wideband ASC is non-trivial due to the need for SNS, digital reconstruction, channelization and characterization algorithms and their integration. {\color{black}To characterize the carrier frequency and DOA, \cite{cascade,anil_sir,isj} are the state-of-the-art methods which use MWC, MCS and FRI based joint estimation of the wideband spectrum.} To the best of our knowledge, \cite{sns_mc1,sns_mc2,sns_mc3,cssp} are the only AMC work which consider ASCW. Their limitations are: 1)~They do not consider state-of-the-art SNS approaches, 2)~Need tunable channelizers to convert wideband spectrum to multiple narrowband signals for characterization using narrowband methods, and 3)~Complex signal conditioning between reconstruction and characterization. We aim to overcome these drawbacks. 
	
 	\section{Proposed ASCW Architecture}
 	
 	In this section, we briefly present the proposed end-to-end architecture for ASCW and elaborately discuss the  architecture in Section~IV and V. We begin with the assumed received wideband signal model.
 	
 	\begin{figure*}[!t]
 	\vspace{-0.2cm}
			\centering
			\includegraphics[scale=0.7]{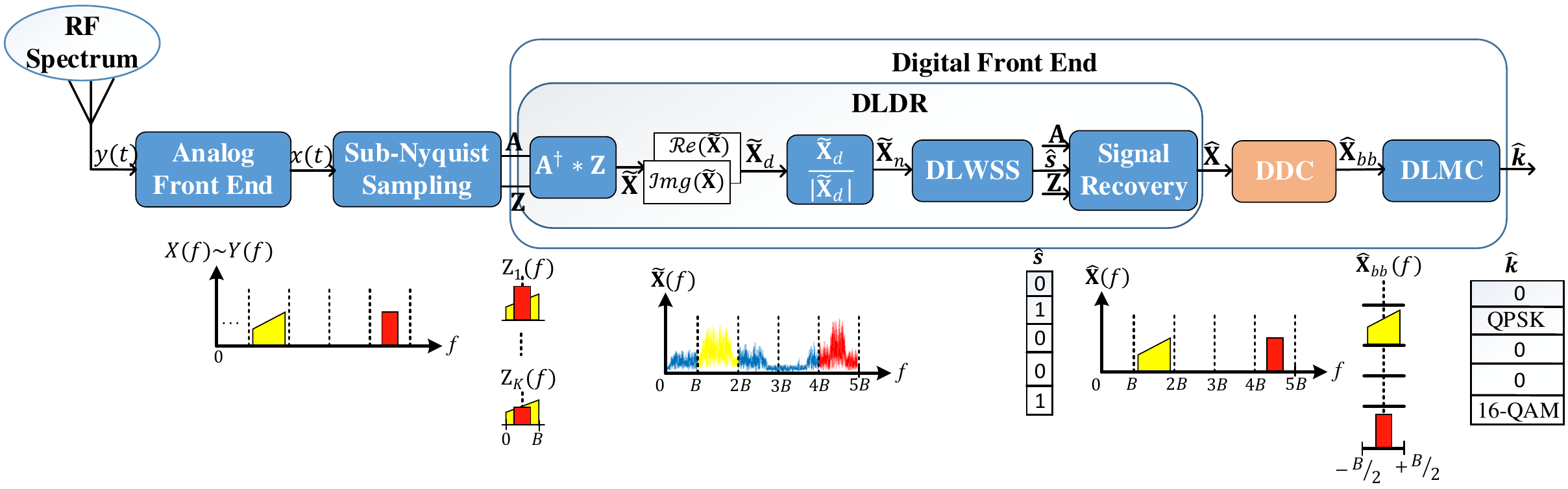}
			\vspace{-0.4cm}
			\caption{Proposed deep learning based architecture for end-to-end ASCW.}
			\vspace{-0.4cm}
			\label{BD}	
	\end{figure*}
	
			\vspace{-0.1cm}
	\subsection{Signal Model}
	Consider a wideband spectrum consisting of a multiple disjoint (i.e. having distinct central frequency) narrowband signals of maximum possible bandwidth, $B~Hz$. Mathematically, the received wideband signal, $ y(t) $ can be modeled as
	\begin{equation}
	\vspace{-0.05cm}
	y(t) = \sum_{p=1}^P h_p(t)*a_p(t) e^{j2\pi f_p t} + \eta(t)
	\vspace{-0.05cm}
	\end{equation}
	where $P$ is the maximum possible number of narrowband signals in $x(t)$, $ a_p(t) $ is the $p^{th}$ modulated narrowband signal of carrier frequency $ f_p $, $ h_p(t) $ is the channel response faced by the $ p^{th} $ signal, $\eta(t)$ is additive white Gaussian noise (AWGN) and $ * $ is a convolution operator. The modulated narrowband signal, $ a_p(t) $ can be represented as
	\begin{equation}
	a_p(t) = \sum_{l=1}^{L}g(t-lT_s)b(l)
	\end{equation}
	where 
	$g(t)$ is the impulse response of a root raised cosine (RRC) pulse shaping filter, 
	$T_s$ is the symbol period, $ b $ is the modulated symbol and $L$ is the length of symbol sequence. Similar to \cite{mcs,mwc,wcl}, we made the following realistic assumptions on the wideband signal:
	
	\begin{enumerate}
		\item The wideband spectrum, $Y(f)$, of Nyquist frequency, $f_{nyq}$ is divided into $N$ frequency bands of bandwidth, $B = \frac{f_{nyq}}{N}$.
		\item The bandwidth of a narrowband signal $a_p(t)$ does not exceed $B~Hz$.
	\end{enumerate}
	
	 For the ease of understanding, the frequently used notations are summarized in Table~\ref{tab_symbol}.
	
	\begin{table}[htbp]
	\vspace{-0.2cm}
		\caption{Notations and their Definitions}
		\label{tab_symbol}
		\vspace{-0.2cm}
		\renewcommand{\arraystretch}{1.3}
		\resizebox{\linewidth}{!}{
			\begin{tabular}{l|l}
				\hline
				\textbf{Notation} & \textbf{Definitions}  \\
				\hline
				$y(t)$ & Received RF wideband signal\\
				\hline
				$x(t)$ & Output wideband signal of AFE\\
				\hline
				$P$ & Maximum number of active transmissions in $x(t)$\\
				\hline
				$a_p(t)$ & $p^{th}$ active transmission in $x(t)$\\
				\hline
				$f_p$ & Carrier frequency of $a_p(t)$ active transmission\\
				\hline
				$h_p(t)$ & Channel response faced by $a_p(t)$ \\
				\hline
				$g(t)$ & Impulse response of RRC filter \\
				\hline
				$b$ & Modulated symbol\\
				\hline
				$T_s$ & Symbol period\\
				\hline
				$L$ & Number of modulated symbols for Dataset, $\textbf{D}_{\textbf{NMC}}$\\
				\hline
				$M$ & Number of modulation schemes to be classified \\
				\hline
				$N$ & Number of frequency bands in $x(t)$\\
				\hline 
				$B$ & Maximum bandwidth of an active transmission\\
				\hline
				$f_{nyq}$ & Nyquist frequency of $x(t)$ \\
				\hline
				$z[n]$ & Sub-Nyquist samples \\
				\hline 
				$Q$ & Number of snapshots produced by every ADCs\\
				\hline 
				$K$ & Number of ADCs used for SNS $(K<<N)$\\
				\hline
				$W$ & Number of observations used for training DL models\\
				\hline
				$\textbf{A}$ & Sensing Matrix of dimension $K\times N$\\
				\hline
				$\textbf{Z}$ & A $K\times Q$ matrix of DTFT of $z[n]$\\
				\hline
				$\textbf{X}(f)$ & A $N\times Q$ matrix containing FT of $N$ frequency bands\\
				\hline
				$\tilde{\textbf{X}}(f)$ & Pseudo-reconstruction of $\textbf{X}(f)$\\
				\hline
				$\hat{\textbf{X}}(f)$ & Reconstructed wideband signal\\
				\hline
				$\hat{\textbf{X}}_{bb}(f)$ & Baseband converted wideband signal, $\hat{\textbf{X}}(f)$\\
				\hline
				$\hat{\textbf{s}}$ & Estimated occupancy status of $N$ frequency bands\\
				\hline
				$\hat{\textbf{k}}$ & Estimated modulation schemes of N frequency bands \\
				\hline
			\end{tabular}}
			\vspace{-0.2cm}
		\end{table} 
		 
		\subsection{ASCW Architecture}
		\label{proposed_WSS_AMC}
		The proposed ASCW architecture, shown in Fig.~\ref{BD}, can be divided into three sections: 1)~AFE for analog signal conditioning, 2) SNS for digitization, and 3) DFE comprising of the proposed DL based reconstruction and characterization.
		
		
		The main task of the AFE is to perform impedance matching, low noise amplification and equalization on the received RF wideband signal, $y(t)$ \cite{mwc_hw}. For simplicity of analysis, we assume, the output, $x(t)$, of AFE is approximately the same as $y(t)$. 
        Next step is SNS based digitization using multiple low-speed ADCs. As discussed in the Section~\ref{Lit_Review}, various SNS architectures like MCS and MWC, which can digitize  contiguous wideband spectrum and FRI based SNS for non-contiguous spectrum, can be used. The discrete time Fourier transform (DTFT) of sub-Nyquist samples, $z[n]$, obtained from SNS can be represented as 
		\begin{equation}
		\label{CS}
		\textbf{Z}(e^{j2\pi f/B}) = \textbf{A} \textbf{X}(f) \qquad\forall~f\in \left[0,B \right] 	
		\end{equation}
		\noindent where $\textbf{A}$ is a $K \times N$ sensing matrix corresponding to the used SNS architecture, $\textbf{X}(f)$ is a $ N \times Q $ matrix containing Fourier transform (FT) of $ N $ frequency sub-bands. Here, $K$ is the number of ADCs used in the SNS and $N$ is the number of sensed frequency bands which in case of the contiguous sensing is same as the total number of bands in $x(t)$ and $Q$ is the number of samples from each ADC. Since the sampling rate of each ADC is $B << f_{nyq}$, all $N$ frequency bands get aliased at the baseband i.e. in the frequency range of $[0,B]$ as shown in Fig.~\ref{BD}.
		
		The aliased sub-Nyquist samples, $\textbf{Z}$, along with the SNS specific sensing matrix, $\textbf{A}$ are passed to the proposed DL based ASCW which identifies the vacant bands along with the modulation scheme (BPSK, QPSK, 16-QAM, 64-QAM, 128-QAM, 256-QAM and 8-PAM) of busy bands. It consists of three stages: 1) DL based digital reconstruction (DLDR) and 2) (Optional) Digital down conversion (DDC) and signal conditioning, and 3) DL based modulation classification (DLMC). Various operations in three stages are shown in Fig.~\ref{BD} and described using Algorithm 1.
		
	 The sub-Nyquist samples, $\textbf{Z}$ and sensing matrix, $\textbf{A}$ are inputs (line~1) while the occupancy status, $\hat{\textbf{s}}$  and identified modulation schemes, $\hat{\textbf{k}}$ of busy bands are the outputs (line~2).
		The inputs, $\textbf{Z}$ and $\textbf{A}$ are first processed to obtain a pseudo-reconstructed signal, $\tilde{\textbf{X}}$, (line~3), given as 
		\begin{equation}
		\tilde{\textbf{X}} = \textbf{{A}}^{\dagger}{\textbf{Z}}
		\end{equation}
		
		\setlength{\textfloatsep}{0pt}
			\begin{algorithm}[!t]
    \caption{Proposed ASCW}
			
			\begin{algorithmic}[1]
				
				\State Input: $ \textbf{A}, \textbf{Z}$
				\State Output: $\hat{\textbf{s}}, \hat{\textbf{k}}$
    \State $\tilde{\textbf{X}} \gets$ $(\textbf{A}^\dagger \textbf{Z})$,
    \State  $\tilde{\textbf{X}}_d \gets$ Concatenate$(\tilde{\textbf{X}}_{real},\tilde{\textbf{X}}_{img})$
    \State  $\tilde{\textbf{X}}_{n} \gets$ Normalize$(\tilde{\textbf{X}}_d)$
    \State  $\theta_{ss} \gets$ DLWSS$(\tilde{\textbf{X}}_{n},\textbf{s})$ \Comment{{\color{black}Training mode}}
    \State  $\hat{\textbf{s}} \gets DLWSS(\tilde{\textbf{X}}_{n},\theta_{ss})$ \Comment{{\color{black}Inference mode}}
      
    \State $\textbf{A}_{new}\gets$ Select columns of $\textbf{A}$ corresponding to busy bands
      
    \State Determine ${\hat{\textbf{X}}}$ according to Eq.~\ref{Xrec}
    \State ${\hat{\textbf{X}}_{bb}} \gets$ DDC and Filtering of ${\hat{\textbf{X}}}$ \Comment{{\color{black}Optional Step}}
    \State $\theta_{c} \gets$ DLMC$(\hat{\textbf{X}}_{bb},\textbf{k})$ \Comment{{\color{black}Training mode}}
    \State $\hat{\textbf{k}} \gets$ DLMC$(\hat{\textbf{X}}_{bb},\theta_{c})$ \Comment{{\color{black}Inference mode}}
    			\end{algorithmic}
		\end{algorithm}

\begin{figure}[!b]
				\centering
				{\includegraphics[scale=0.8]{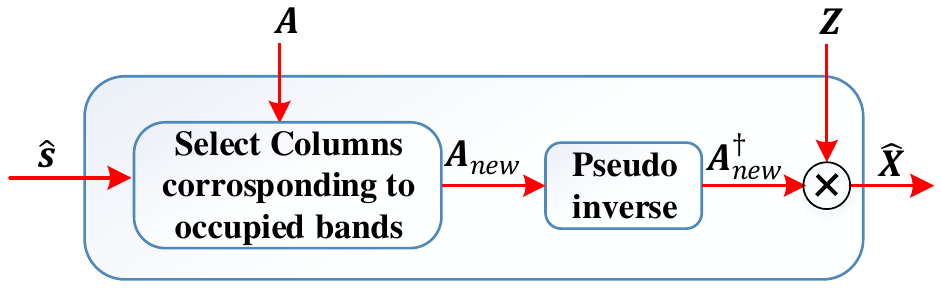}}
				\vspace{-0.2cm}
				\caption{Wideband signal recovery in the DLDR block.}
				\label{Rec}
				\vspace{-0.2cm}
			\end{figure}

		Since $\tilde{\textbf{X}}$ is a complex signal  of dimension $N\times Q$ and can not be feed to the DL model directly, it is reshaped to a higher dimensional matrix, $\tilde{\textbf{X}}_d$ of size $N \times Q \times 2$ (line~4) and the third dimension of size $2$ represents the real and imaginary values of $\tilde{\textbf{X}}$. For the faster convergence of the training process, the higher dimensional pseudo-reconstructed matrix, $\tilde{\textbf{X}}_d$ is normalized between  $[0,1]$ (line~5). The normalized matrix, represented by $\tilde{\textbf{X}}_{n}$, is fed to DL based wideband spectrum sensing (DLWSS) block.  The DLWSS is based on a convolution neural network (referred as $CNN_{SS}$) and its architecture along with the ablation study is discussed later in Section~\ref{dlwss}. The output of the DLWSS is $\hat{\textbf{s}}$ which contains the status of each digitized frequency band. Note that $\hat{s}_n = 0~(or~1)$ denotes that $n^{th}$ band is vacant (or occupied, respectively). The DL models work in two modes: 1) Offline training mode and 2) Online inference mode (i.e. testing mode). Offline training mode provides the learned network parameters, $\theta_{ss}$ (line~6) which are then used to determine  occupancy status, $\hat{\textbf{s}}$ (line~7) of bands in the digitized spectrum, $\tilde{\textbf{X}}_n$.
		Since the reconstruction noise is very high in the pseudo-reconstructed signal, $\tilde{\textbf{X}}$, we perform the signal reconstruction (line~8-9) and corresponding architecture is shown in Fig.~\ref{Rec}. Here, we take band occupancy status vector, $\hat{\textbf{s}}$, determined by the DLWSS block, the sub-Nyquist samples, $\textbf{Z}$ and the sensing matrix $\textbf{A}$ as inputs. First, we generate a new sensing matrix $\textbf{A}_{new}$ by selecting the columns of $\textbf{A}$ which corresponds to the busy frequency bands (line~8) followed by the  reconstruction of wideband signal (line~9). Mathematically, the reconstruction step can be written as 
			\begin{equation}
			\hat{\textbf{X}}[n] =  \left\{
			\begin{IEEEeqnarraybox}[][c]{l?s} 
			\IEEEstrut 
			\textbf{A}_{new}^\dagger \textbf{Z} & $\forall ~\hat{s}(n) = 1$, \\ 
			0 & otherwise
			\IEEEstrut 
			\end{IEEEeqnarraybox} 
			\right.
			\label{Xrec}
			\end{equation} 
			where $\hat{\textbf{X}}[n]$ is the reconstructed wideband signal for $n^{th}$  frequency band, $\textbf{A}_{new}^\dagger$ is the pseudo-inverse of new sensing matrix and $\textbf{Z}$ is the DTFT of sub-Nyquist samples.


  After reconstruction, the next step is the modulation classification of all the active bands present in the digitized wideband spectrum. As discussed before, we propose simultaneous multi-band classification as opposed to sequential single band classification in existing works \cite{ref3,ref4,sns_mc1,sns_mc2,sns_mc3}. In this direction, we consider two scenarios:

    
    
    \subsubsection{Scenario~1}
    In the first scenario, we employ DDC to bring each occupied signal at the baseband before the modulation classification. The output of this block, as shown in Fig.~\ref{BD}, is represented as $\hat{\textbf{X}}_{bb}$ and is of size $N\times L$ where $L$ is the number of modulated symbols. Depending on the requirement, it is represented into one of the two forms: 1) Amplitude and phase, and 2) Real and imaginary. This is followed by the proposed narrowband DL based modulation classifier (NDLMC) that classify all the bands simultaneously as compared to sequential single band classification in existing approaches \cite{ref3,ref4,sns_mc1,sns_mc2,sns_mc3}. Furthermore, the proposed approach is more sophisticated than a Velcro approach of stacking multiple classifiers in parallel, and we explore a new formulation of cross-entropy loss function to simultaneously classify multiple bands in wideband signal.
    
    
    \subsubsection{Scenario~2}
    Since the DDC of each occupied band incurs significant computational complexity, we directly characterize the reconstructed wideband signal, $\hat{\textbf{X}}$ via new wideband DLMC (WDLMC) discussed later in Section V. Another benefit of WDLMC is its architecture similarity with DLWSS making the complete architecture a good candidate for reconfigurable platforms such as Zynq SoC. 
    

		

		\section{Proposed DLWSS Architecture}
		\label{dlwss}
\begin{figure}[!t]
		\centering
		{\includegraphics[scale=0.7]{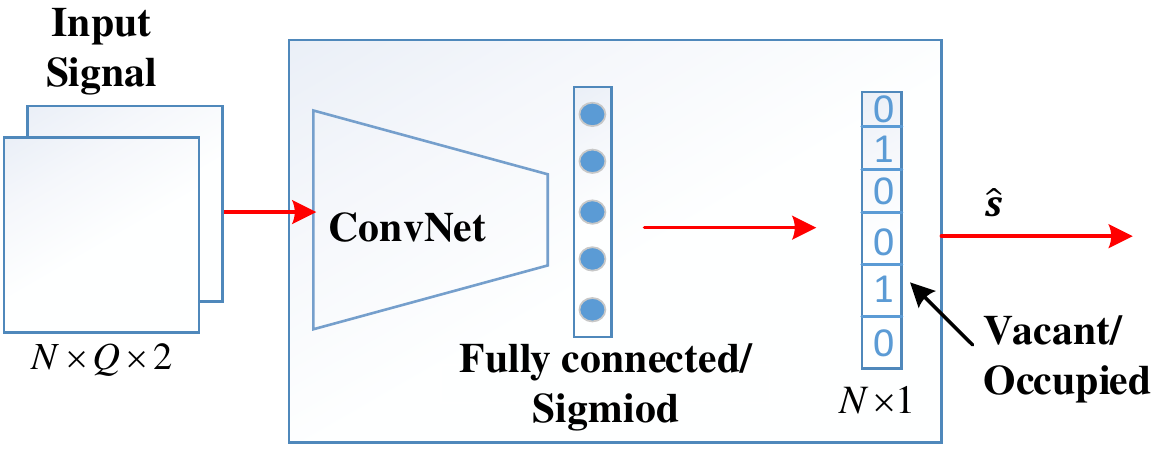}}
		\caption{CNN architecture for DLWSS. }
		\label{DLSS}
	\end{figure}

		The DLWSS aims to determine the occupancy status, $\hat{\textbf{s}}$, of the wideband spectrum for subsequent reconstruction and characterization. 
        The proposed DLWSS is based on the convolutional neural network (CNN)\footnote{The CNN is preferred over LSTM based on in-depth study and comparison for a wide variety of DLWSS datasets. The details are skipped due to limited space constraints.} and its architecture is shown in Fig.~\ref{DLSS}. As discussed in Section~\ref{proposed_WSS_AMC}, we first perform offline training to learn the network parameters, $\theta_{ss}$ followed by the testing of real-time pseudo-reconstructed signal in the inference/testing mode. 
		Algorithm~2 shows the offline training process with the dataset, $\textbf{D}_{WSS} = \{(\tilde{\textbf{X}}_{n,1},\textbf{s}_1), (\tilde{\textbf{X}}_{n,2},\textbf{s}_2), ..... , (\tilde{\textbf{X}}_{n,W},\textbf{s}_W)\}$ where $W$ denotes the number of observations (or examples) over which training is performed, $\tilde{\textbf{X}}_{n,w} \in \mathfrak{R}^{N\times Q \times 2}$ is the $w^{th}$ normalized and pseudo-reconstructed signal, $\tilde{\textbf{X}}_{n}$ and $\textbf{s}_w\in \{0,1\}^{N\times 1}$ is the label of $w^{th}$ observation (or example) indicating the actual occupancy status 
		of all $N$ frequency bands of $\tilde{\textbf{X}}_{n,w}$ (line~5-7).
\vspace{1cm}
	
\begin{algorithm}[!t]
			\caption{DLWSS Training Mode}
			\begin{algorithmic}[1]
				\State Input: Dataset$=\{\tilde{\textbf{X}}_{n,w},\textbf{s}_w\}\forall w\in\{1,W\},~\eta,~t=0$
				\State Output: $\theta_{ss}$
				\State Initialize: $\theta_{ss} = \mathcal{N}(0,1)$
                \While{{\color{black}not converge}}
                \State $t = t+1$
				\State Sample  batch of data-points $\{\tilde{\textbf{X}}_{n,w},\textbf{s}_w\}$
				\State  $\hat{\textbf{s}}\gets CNN_{SS}(\tilde{\textbf{X}}_{n,w},\theta_{ss})$
				\State Calculate $L_{BCE}$ as per Eq.~\ref{train_loss}
				\State Update $\theta_{ss}$ as per Eq.~\ref{update_ss}
				\EndWhile
			\end{algorithmic}
		    \end{algorithm}
        The training process involves the minimization of a loss function, which is a measure of inconsistency between the predicted and actual label. 
		Since more than one frequency band can be busy in a wideband spectrum, the problem is formulated as a multi-label binary classification with binary cross entropy as the training loss function. It is calculated as
		\begin{equation}
		L_{BCE}(\hat{\textbf{s}},\textbf{s}) = - \sum_{n=1}^{N}(s(n) \log \hat{s}(n) + (1-s(n))\log(1-\hat{s}(n)) )
		\label{train_loss}
		\end{equation}
		where $s(n)\in \textbf{s}$ and $\hat{s}(n)\in \hat{\textbf{s}}$ is the  actual and predicted  occupancy status of $n^{th}$ frequency band (line~8).
		Furthermore, the learnable network parameters, $\theta_{ss}$, are optimized using a stochastic gradient descent algorithm such that the training loss, $L_{BCE}$, is minimized. Mathematically, it is represented as
		\begin{equation}
		    \theta_{ss} = \theta_{ss}-\eta \nabla_{\theta_{ss}}L_{BCE} 
		    \label{update_ss}
		\end{equation}
		where $\eta$ is the learning rate.
        Here, the loss gradients, $\nabla_{\theta_{ss}}L_{BCE}$ are backpropagated and used to update the learnable network parameters at each iteration (line~9). This process is iterative and repeated until the saturation of the validation loss (i.e. till the model does not converge). The final output is the optimized parameters, $\theta_{ss}$.
		After the training mode, the CNN model is used in the inference mode to find the occupancy status of unknown wideband spectrum in real-time.
		

		
\begin{figure}[!h]
				\centering
				\vspace{-0.2cm}
				\includegraphics[scale=0.5]{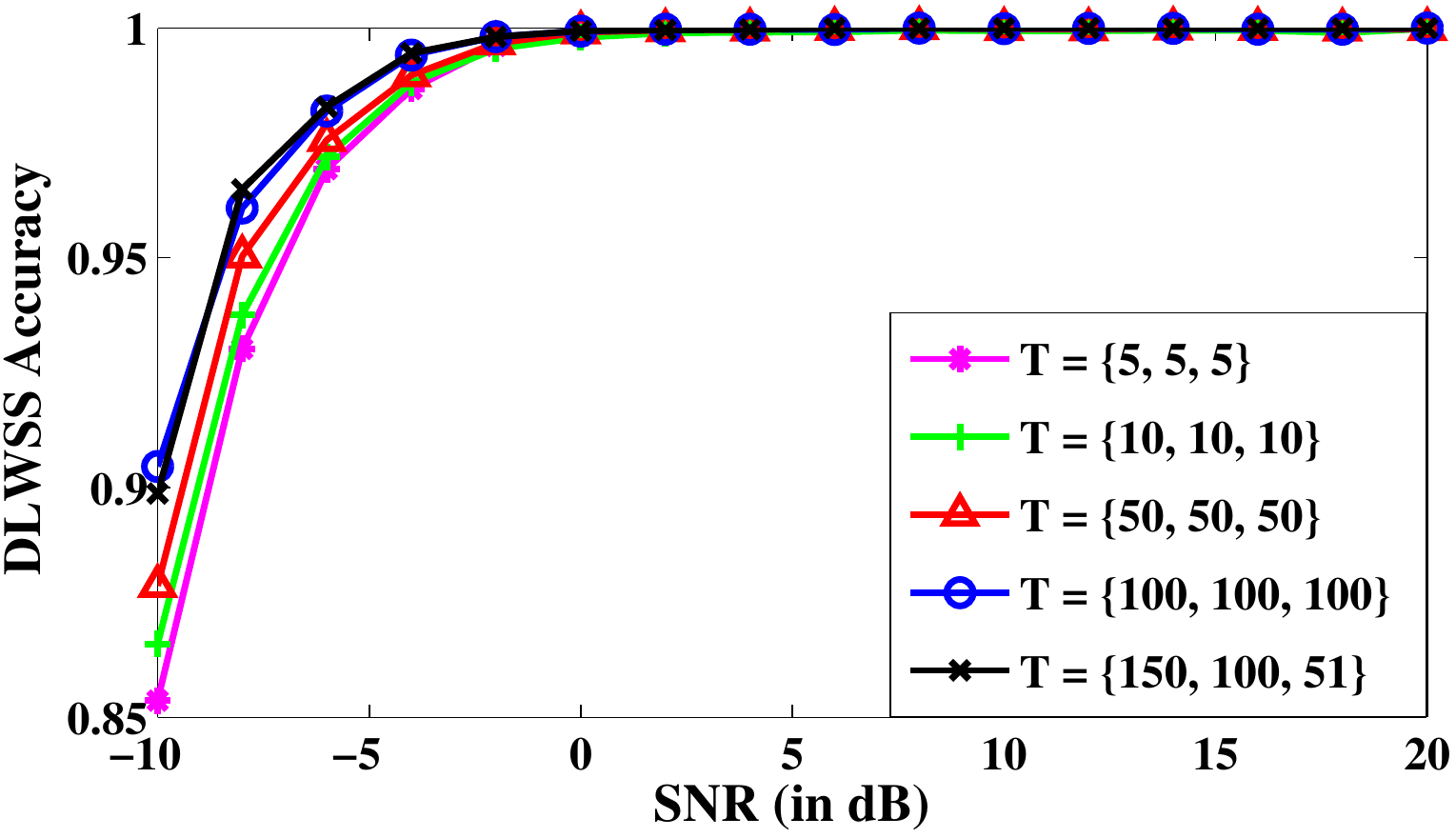}
				\vspace{-0.2cm}
				\caption{Classification accuracy of CNN based DLWSS for different values of n-taps.}
				\label{abl_DLWSS}
			\end{figure}

		The ablation study of DLWSS CNN model is shown in Fig.~\ref{abl_DLWSS}.
		The experiments are performed for different network depths and filter setting. The filters used are of the form 1$\times$ n-taps where n-taps denotes the width of the convolution filter in all layers. To perform the ablation for filter size, we fix the number of filters to $64$ in all layers. We observe that filters with a larger width perform better as compared to those with a smaller width and saturates when the width is increased further. The same has been shown in Fig.~\ref{abl_DLWSS} for three-layer CNN. Here, $T$ tells the n-taps values for the three layers of CNN.
		The best classification accuracy is obtained when $T=\{150, 100, 51\}$. Furthermore, we then fix the filter sizes to the best case  and vary the number of filters in each layer. The best performing architecture is shown in Table~\ref{CNN_para}. The same architecture has been selected for the rest of the discussion. 

		\begin{table}[!h]
			\begin{center}
				\vspace{-0.2cm}
				\caption{CNN Architecture for the proposed DLWSS.}
				\label{CNN_para}
					\vspace{-0.25cm}
				\renewcommand{\arraystretch}{1.2}
				\resizebox{\linewidth}{!}{
					\begin{tabular}{c|c|c|c}
						{\textbf{Layers}} & \textbf{Filter Size} & {\textbf{Number of Filters }} & {\textbf{Output Dimension}}\\
						\hline
						Input & $-$ & $-$ & $N\times Q \times 2$\\
						Conv/relu & 1x150 & 256 & $N\times 150\times 256$\\ 
						Conv/relu & 1x100 & 128 & $N\times 51\times128$\\ 
						Conv/relu & 1x51 & 64 & $N\times 1\times 64$\\ 
						Custom pool & $-$ & $-$ & $N\times 1\times 64$\\
						FC/sigmoid & $-$ & $-$ & $N$
					\end{tabular}}
				\end{center}
			\end{table}
			
\vspace{5cm}
		\section{Proposed DLMC Architecture}
		As discussed in Section III, to perform DLMC, we consider two scenarios: 1) NDLMC, and 2) WDMLC. We present the corresponding architectures below.
		
		\subsection{NDLMC Architecture}
		As shown in Fig.~\ref{BD}, the output of DLDR block, $\hat{\textbf{X}}$, is processed via DDC to obtain $\hat{\textbf{X}}_{bb}$ of size $N\times L$. To perform AMC on this processed baseband spectrum, three types of datasets have been considered in the literature: 1) IQ (in-phase and quadrature-phase) samples \cite{ref4}, 2) Amplitude-Phase (AP) samples \cite{ref3} and 3) Constellation diagram images of modulation schemes \cite{constellation}. Since received wideband spectrum is represented using complex samples, IQ and AP datasets are preferred and readily available in the wireless receiver without the need of additional processing in physical layer compared to the constellation image based processing. Hence, we restrict the discussion to IQ and AP samples of $\hat{\textbf{X}}_{bb}$. Furthermore, our models are designed considering the various studies which show that the CNN and LSTM models are more suited for IQ and AP samples, respectively, \cite{ref1,ref2,ref3,ref4,ref5,ref6,comsnet}. 
		
		
			Algorithm~3 shows the steps involved in the training of NDLMC. Inputs to this classifier are $\hat{\textbf{X}}_{bb}$ and the labels of modulation schemes, $\textbf{k}_w$.
		    Since $M$ modulation schemes are considered for the classification for each of the $N$ frequency bands, the output of the classifier is a vector, $\hat{\textbf{h}}$ of un-normalized log probabilities and has the size of $1\times M$. The values in vector $\hat{\textbf{h}}$ are converted to probabilities by applying a softmax activation function which for a particular band is calculated as
			\begin{equation}
			\hat{p_i} = \mathbf{softmax}(\hat{h}_i) = \frac{exp(\hat{h_i})}{\sum_{j}exp(\hat{h}_j))}
			\end{equation}
			\vspace{0.1cm}
			
\setlength{\textfloatsep}{1pt}
		  \begin{algorithm}[!b]
			\caption{NDLMC Training Mode}
			\begin{algorithmic}[1]
				\State Input: Dataset$=\{\hat{\textbf{X}}_{bb,w},\textbf{k}_w\}\forall w\in\{1,W\},~\eta,~t=0$
				\State Output: $\theta_{c}$
				\State Initialize: $\theta_{c} = \mathcal{N}(0,1)$
                \While{{\color{black}not converge}}
                \State $t = t+1$
				\State Sample  batch of data-points $\{\hat{\textbf{X}}_{bb,w},\textbf{k}_w\}$
				\State  $\hat{\textbf{k}}\gets $ Classifier$(\hat{\textbf{X}}_{bb,w},\theta_{c})$
				\State Calculate $L_{c}$ as per Eq.~\ref{train_loss_mc}
				\State Update $\theta_{c}$ as per Eq.~\ref{update_ss_mc}
				\EndWhile
			\end{algorithmic}
		\end{algorithm}
			where $\hat{p_i}$ is the predicted probability of $i^{th}$ modulation scheme for a frequency band. This gives an output vector of size $N\times M$ (line~5-7). Next, similar to the CNN modeling for DLWSS, to optimise the network learnable parameters for the modulation scheme classifiers, we use a stochastic gradient descent algorithm to minimize the training loss. As the training loss for a particular frequency band depends on the status of the band, we define it as the categorical cross entropy if the band is busy and zero if the band is vacant. Mathematically, loss per band can be defined as
			\begin{equation}
			L_{n} = \left\{ \,
			\begin{IEEEeqnarraybox}[][c]{l?s}
			\IEEEstrut
			-\sum_{i}p_i log(\hat{p_i}) & if $\hat{s}_n = 1$, \\
			0 & if $\hat{s}_n = 0$
			\IEEEstrut
			\end{IEEEeqnarraybox}
			\right.
			\end{equation}
			where $L_{n}$ is the loss of $n^{th}$ frequency band, $\hat{s}_n$ is the occupancy status of the $n^{th}$ band and ${p_i}$ is the actual probability of the $i^{th}$ modulation scheme for a particular frequency band.
			
			To determine the complete loss function (line~8), we concatenate the band status vector, $\hat{\textbf{s}}$ with the detected modulation scheme vector, $\hat{\textbf{k}}$ and it can be precisely expressed as
			\begin{equation}
			L_c = \sum_{n=1}^{N} -\hat{s}_n \left(\sum_{i}p_i \log\hat{p_{\textcolor{black}i}}\right)
			\label{train_loss_mc}
			\end{equation}
			where $\hat{s}_n$ is the status (i.e. $0$ for vacant and $1$ for busy) of the $n^{th}$ frequency band. Now, similar to DLWSS, based on the loss gradient, the network learnable parameter, $\theta_c$, is updated as (line~9)
			\begin{equation}
		    \theta_{c} = \theta_{c}-\eta \nabla_{\theta_{c}}L_{c} 
		    \label{update_ss_mc}
		\end{equation}

		
						\begin{table}[!b]
										\begin{center}
											\caption{$CNN_{Baseline}$ Architecture for NDLMC.}
											\vspace{-0.25cm}
											\label{CNN_IQ}
							\renewcommand{\arraystretch}{1.2}
											\resizebox{\linewidth}{!}{
											\begin{tabular}{c|c|c|c}
												{\textbf{Layers}} & \textbf{Filter Size} & {\textbf{Number of Filters }} & {\textbf{Output Dimension}}\\
												\hline
												Input & - & - & $N\times L \times 2$\\
												Conv/relu & $1\times 3$ & 64 & $N\times L\times 64$\\ 
												Conv/relu & $1\times 3$ & 64 & $N\times L\times 64$\\ 
												Conv & $1\times 1$ & $M+1$ & $N\times L\times (M+1)$\\ 
												Custom~pool/softmax & - & - & $N\times (M+1)$
											\end{tabular}}
										\end{center}
											\vspace{-0.2cm}
									\end{table}
									
				\renewcommand{\thetable}{V}	
			\begin{table*}[!b]
				\begin{center}
					\vspace{-0.2cm}
					\caption{Modulation classification accuracy of various variants of CNN model for NDLMC and WDLMC.} 
                    \label{DL_class}
					\vspace{-0.35cm}
					\renewcommand{\arraystretch}{1.2}
					\resizebox{\linewidth}{!}{
						\begin{tabular}{|c|c|c|c|c|c|c|c|c|c|c|c|c|c|c|c|c|c|c|c|c|}
							\hline
							\textbf{DLMC} & \multirow{2}{*}{\textbf{Classifiers}} & \multicolumn{5}{c|}{\textbf{AWGN}} & \multicolumn{5}{c|}{\textbf{Rician Fading with Doppler}} & \multicolumn{5}{c|}{\textbf{Rayleigh Fading with Doppler}}\\
							\cline{3-17}
							\textbf{Methods}  & & $-10~dB$ & $-6~dB$ & $0~dB$ & $4~dB$ & $14~dB$ & $-10~dB$ & $-6~dB$ & $0~dB$ & $4~dB$ & $14~dB$ & $-10~dB$ & $-6~dB$ & $0~dB$ & $4~dB$ & $14~dB$  \\ 				
							\hline
							\multirow{6}{*}{\textbf{NDLMC}}&\textbf{Baseline} &  56.4 & 72.3 & 85.1 & 89.2 & 96.4 &  45.4 & 59.3 & 72.6 & 74.7 & 77.1 & 43.2 & 58.7 & 71 & 71 & 76.8 \\
							\cline{2-17}
							& \textbf{NiN} & 58.4 & 72.3 & 84.5 & 91.7 & 99.9 & 45.1 & 59.7 & 72.9 & 78.1 & 89.1 &  44.02 & 59.4 & 74.1 & 79.1 & 89.6\\
							\cline{2-17}
							& \textbf{ResNet} & 58.5 & 72.9 & 84.6 & 91.4 & 99.9 & 45.9 & 61.6 & 74.5 & 79.1 & 89.7  & 46.2 & 61.2 & 73.4 & 79 & 89.1\\
							\cline{2-17}
							& \textbf{Densenet} & 57.1 & 72.9 & 85.3 & 90.2 & 99.8 & 46.8 & 61.8 & 75.9 & 79.8 & 90 & 45.8 & 60.9 & 73.5 & 79 & 90 \\
							\cline{2-17}
							& \textbf{Inception-SNS} & \textbf{59} & \textbf{73.7} & \textbf{84.4} & \textbf{91.7} & \textbf{99.8}  & \textbf{47.5} & \textbf{62.3} & \textbf{74.8} & \textbf{79.8} & \textbf{89.4} & \textbf{45.3} & \textbf{61.6} & \textbf{73.4} & \textbf{78.5} & \textbf{90.3} \\
							\cline{2-17}
							& \textbf{Inception-NS} & 68.6 & 81.5 & 90.1 & 97.7 & 100 & 54.6 & 68.3 & 79.5 & 86 & 93.5 & 57.2 & 68.6 & 78.3 & 83.3 & 91.1 \\
							\hline
							
							\multirow{2}{*}{\textbf{WDLMC}} & \textbf{CNN-{SNS}} & 42 & 49.7 & 67.1 & 75.4 & 82.7 & 34.8 & 39.7 & 45.9 & 52.6 & 52 & 34.3 & 37.9 & 42.1 & 46 & 47.7 \\
							\cline{2-17}
							
							& \textbf{CNN-{NS}} & 43.8 & 58.1 & 74.5 & 79.1 & 83.2 & 34.4 & 44.7 & 50.2 & 51.5 & 53.2 &  33.1 & 42.1 & 45.4 & 47.8 & 48.3 \\
							\hline
						\end{tabular}
					}
					\vspace{-0.2cm}
				\end{center}
			\end{table*} 
		Next step is to finalize the architectures of CNN and LSTM based models via detailed ablation study. Such study is important since the proposed architectures are novel in the sense that a multiband signal classification is considered compared to sequential single band classification in existing works \cite{ref3,ref4}.  In the proposed setup, the final outcome is of the form $N \times (M+1)$ comprising of the status of $N$ bands and the modulation schemes of busy bands. 
		
\begin{figure}[!t]
				\centering
					\vspace{-0.2cm}
				\subfloat[]{\includegraphics[scale=0.275]{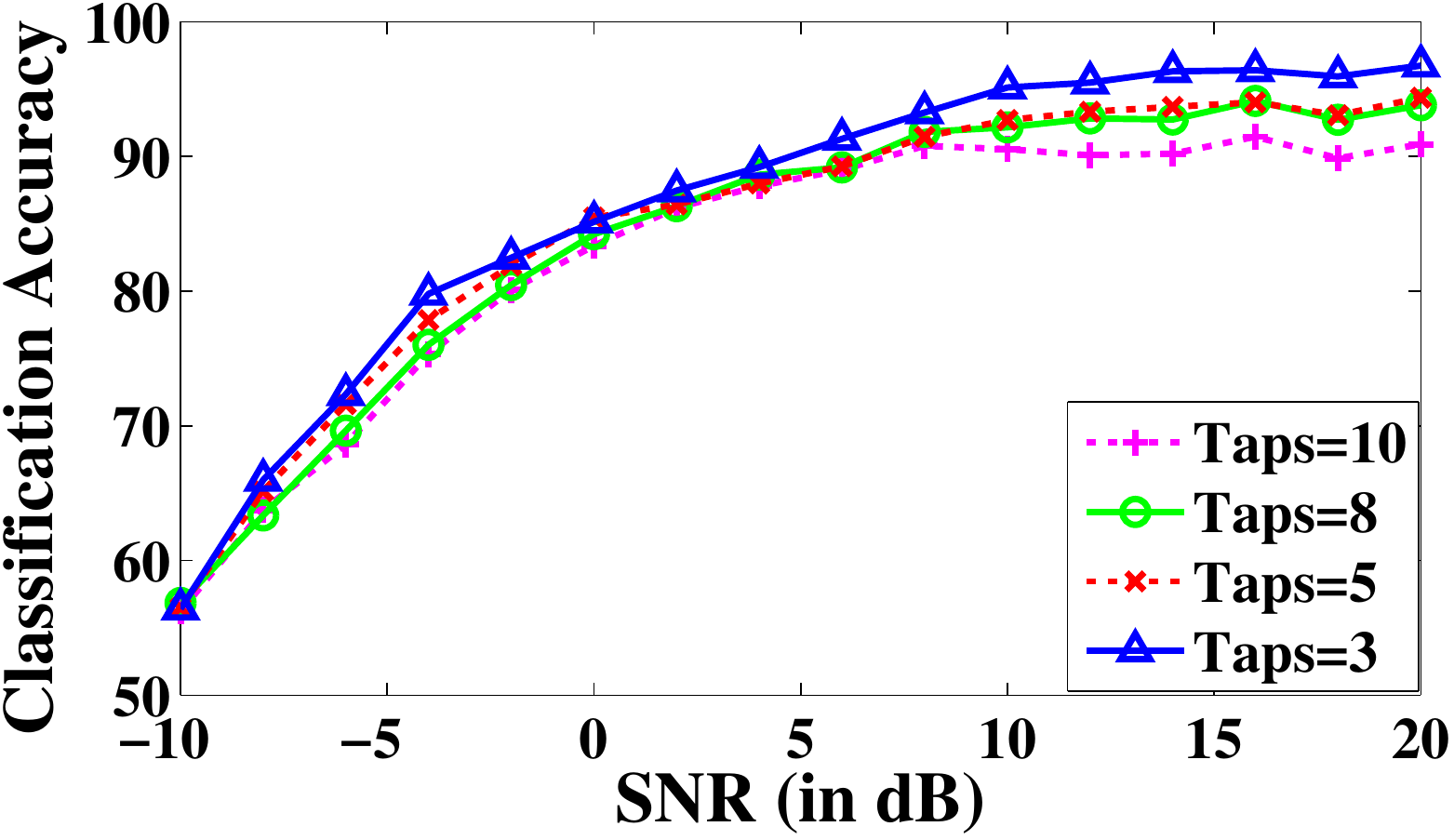}}
				\hspace{0.01cm}
			 \subfloat[]{\includegraphics[scale=0.275]{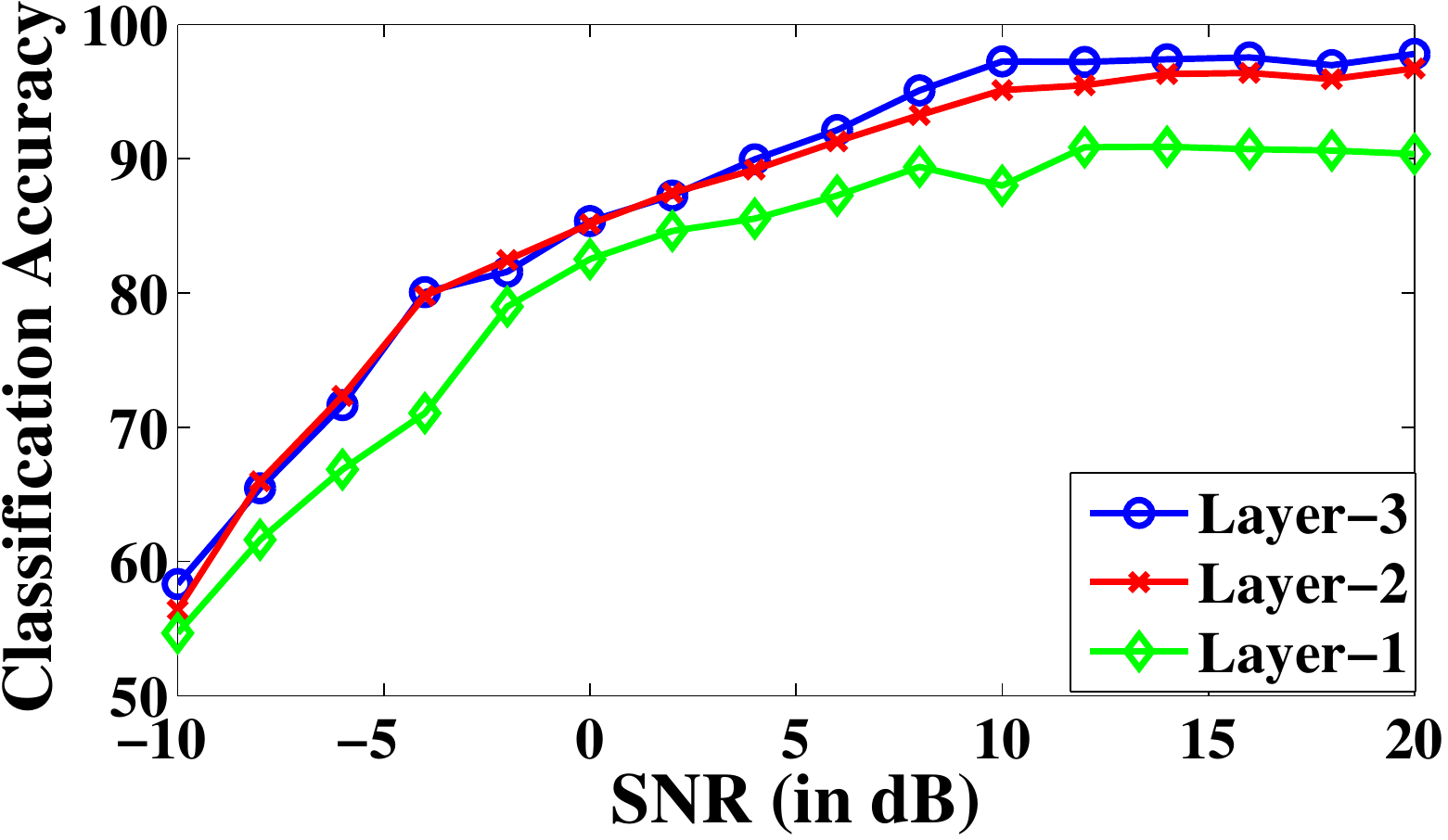}}
			 	\vspace{-0.2cm}
				\caption{Ablation study for $CNN_{Baseline}$ on IQ samples for different values of (a) n-taps, and (b) Layers.}
				\label{Basline_CNN}
				\vspace{-0.25cm}
			\end{figure}

\begin{figure}[!t]
	            \vspace{-0.2cm}
				\centering
				\includegraphics[scale=0.55]{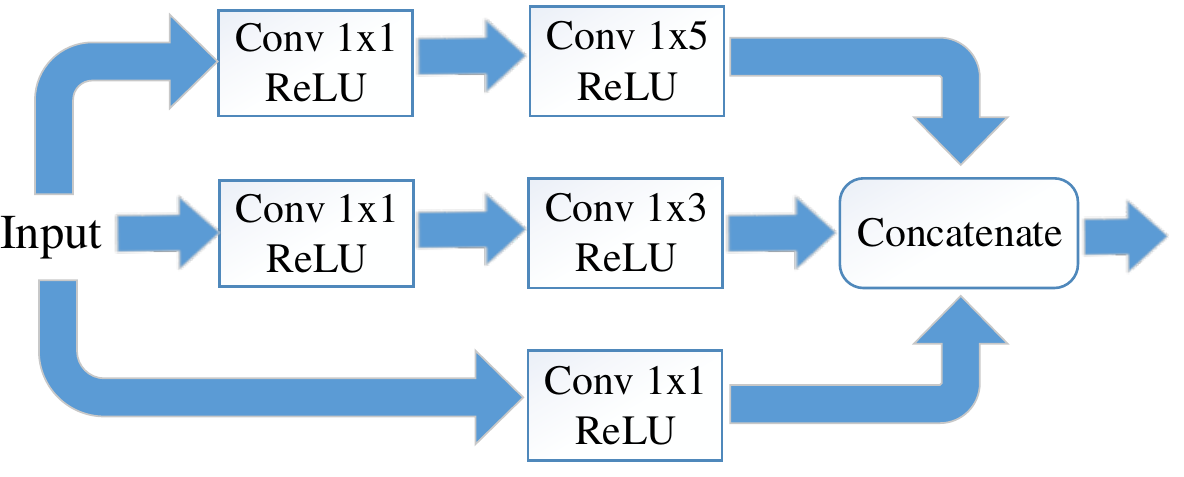}
				\vspace{-0.25cm}
				\caption{Architecture of Inception block.}
				\label{inception}
			\end{figure}

		The baseline CNN (referred as $CNN_{Baseline}$) architecture that we considered is shown in Table~\ref{CNN_IQ}.	We studied the classification performance to decide our baseline model architecture for different sizes of filter (i.e. n-taps), number of filter and depth of the network. Fig.~\ref{Basline_CNN} (a) shows the classification performance for various values of n-taps. We notice that smaller filters (i.e. for n-taps $=3,5$) perform better than the larger filters. This is an important observation as it is not valid in case of WDLMC discussed later in Table~\ref{CNN_raw}.  Thus, we use small filter sizes n-taps $ = 3$ in the $CNN_{Baseline}$. Furthermore, the results obtained by varying the number of filters are very similar to the ones obtained in \cite{ref2}. Thus, we use $64$  number of filters for further analysis as it is efficient from both computation, memory and performance  point  of  view. Also, as shown in Fig.~\ref{Basline_CNN} (b), we observe no significant performance improvements when we increase the number of layers with $1\times$n-taps filters beyond $2$. Hence, the $CNN_{Baseline}$ 
			has two convolution layers with $1\times$n-taps filter size, a convolution layer of filter size $1\times 1$, and a custom average pool/softmax activation layer at the end. The last convolution layer of filter size $1\times 1$, along with the custom pool layer aim to match the dimension of the output label (i.e. $M+1$) and it is performed by averaging the input along the column dimension (i.e. $L$). Note that the softmax activation layer associates the output with the probability of occurrence of every modulation scheme. Furthermore, we consider four variants to enhance the AMC performance of  $CNN_{Baseline}$: 1) Network in Network(NiN) \cite{ref9}, 2) Inception network \cite{ref11}, 3) Residual network \cite{ref10}, and 4) Densenet network \cite{ref12}. The modulation classification accuracy of these variants for different channel conditions, i.e. AWGN, Rayleigh and Rayleigh with a Doppler shift of 10Hz is shown in Table~\ref{DL_class}. It can be observed that Inception offers better performance than other architectures, and hence, it is chosen for the NDLMC task with IQ-samples as input. Note that the performance of NDLMC improves if Nyquist sampling is used. Table~\ref{cnn_models} and Fig.~\ref{inception} show the corresponding architecture of Inception model for NDLMC and Inception block. The proposed LSTM based NDLMC architecture for AP samples along with ablation study and the performance analysis is discussed in Appendix A.



			

			
			
		\renewcommand{\thetable}{IV}	
		\begin{table}[!t]
		\begin{center}
			\caption{{Architecture of Inception model for NDLMC.}}
			\label{cnn_models}\vspace{-0.2cm}
			{
				\begin{tabular}{c|c}
					
				\cline{1-2}
				\textbf{Layer} & \textbf{Output dimension} \\
				\cline{1-2}
				Input & $N\times L \times 2$ \\
				Inception Block & $N\times L\times 192$ \\ 
				Inception Block & $N\times L\times 192$ \\
				$1\times 1$ Conv/Relu & $N\times L\times (M+1)$ \\
                Custom~pool/softmax & $N\times (M+1)$ \\
					\cline{1-2}
				\end{tabular}}
			\end{center}
		\end{table}
		
			
\begin{figure}[!t]
	\vspace{-0.2cm}
				\centering
				\includegraphics[scale=0.525]{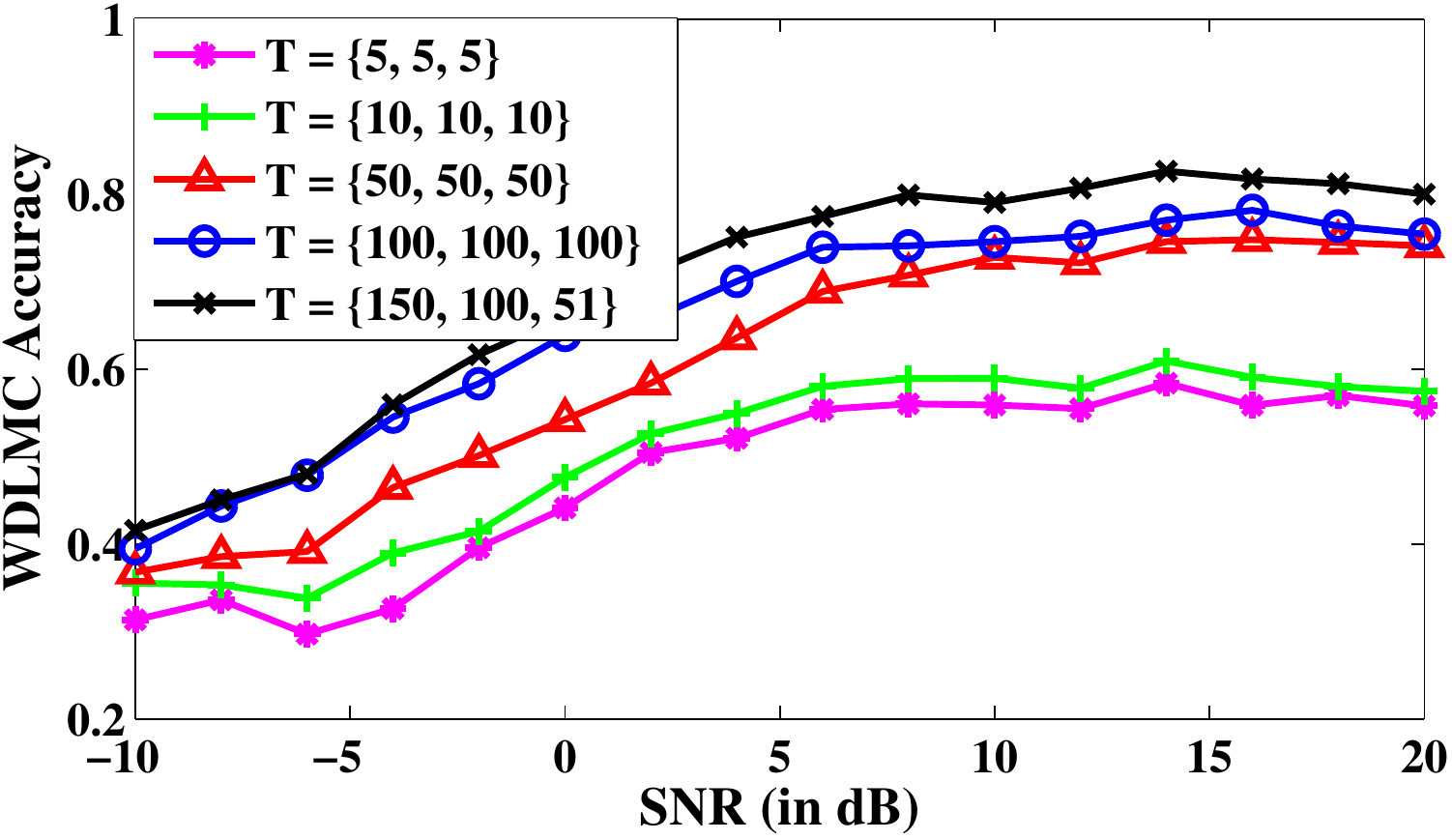}
				\vspace{-0.25cm}
				\caption{Ablation study of CNN based WDLMC architecture for different values of $n$-taps.}
				\label{abl_WDLMC}
			\end{figure}


		\renewcommand{\thetable}{VI}	
			\begin{table}[!t]
				\begin{center}
					\vspace{-0.2cm}
					\caption{CNN Architecture for the proposed WDLMC.}
					\label{CNN_raw}
							\vspace{-0.25cm}	\renewcommand{\arraystretch}{1.3}
								\resizebox{\linewidth}{!}{
									\begin{tabular}{c|c|c|c}
										{\textbf{Layers}} & \textbf{Filter Size} & {\textbf{Number of Filters }} & {\textbf{Output Dimension}}\\
						\hline
						Input &  &  & $N\times Q \times 2$\\
						Conv/ReLu & $1\times150$ & 256 & $N\times 150\times 256$\\ 
						Conv/ReLu & $1\times 100$ & 128 & $N\times51\times128$\\ 
						Conv/ReLu & $1\times 51$ & 64 & $N\times 1\times 64$\\ 
						Conv/ReLU & $1\times 1$ & $M+1$ & $N\times 1\times (M+1)$\\
						Custom pool/softmax & & & $N\times (M+1)$
					\end{tabular}}
				\end{center}
			\end{table}

		\subsection{WDLMC Architecture}
		
		Next, we consider AMC of the wideband spectrum directly from the reconstructed wideband signal, $\hat{\textbf{X}}$ (i.e. raw samples). The training of the WDLMC  follows the same approach as that of NDLMC in algorithm~3 except that $\hat{\textbf{X}}$ is used instead of $\hat{\textbf{X}}_{bb}$ in Line~1. In the end, we have estimated network learnable parameters, $\theta_c$ via steepest descent method (line~8-9). Similar to Section~V.A, we present the CNN based architecture here and LSTM based architecture in Appendix A.
		Fig.~\ref{abl_WDLMC} shows the ablation study for the CNN classifier for WDLMC. The chosen WDLMC architecture is given in Table~\ref{CNN_raw}. Note that the additional convolution layer of filter size $1\time 1$ in the architecture is to match the output of the model to the dimensions of the ground truth label which are $N\times (M+1)$. The performance of WDLMC for different wireless channels is shown in the last rows of Table~\ref{DL_class} along with its performance for Nyquist sampled data. As expected, the performance of WDLMC is poorer than NDLMC, especially for wireless channels with Rayleigh and Rician fading. However, the performance of the proposed architecture for Nyquist and sub-Nyquist sampled data is almost identical for higher SNRs. This indicates that the design of WDLMC architectures for these channels is itself a challenging research problem. This problem is important for architecture perspective as the design of intelligent and reconfigurable physical layer is one of the critical research areas. In this direction, the significant similarity between the proposed WDLMC and DLWSS architecture in terms of number of layers, number of filters and size makes the proposed ASCW transceiver well-suited for realization on the reconfigurable hardware. For example, via dynamic partial reconfiguration capability of Zynq SoC, on-the-fly switch between DLWSS and WDLMC can offer a significant reduction in area and power complexity along with cost benefits due to the reduction in the chip area. Thus, the proposed WDLMC approach is novel state-of-the-art contribution in the design of reconfigurable ASCW. Similar to NDLMC, we also consider the LSTM based WDLMC architecture for completeness of the performance analysis and corresponding details are given in Appendix A.
		
		
					\vspace{-0.25cm}
			\section{Simulation Setup}
			In this section, we discuss the proposed state-of-the-art datasets for training and verification of DLWSS and DLMC. The proposed datasets are generated synthetically using MATLAB and are the only available datasets for SNS based ASCW. Each dataset is keyed with modulation scheme and SNR.
            We consider seven widely used modulation schemes (BPSK, QPSK, 16-QAM, 64-QAM, 128-QAM, 256-QAM, 8-PAM) and SNR ranging from -10dB to 20dB with an interval of 2dB. We consider $N=14$ frequency bands and MCS based SNS for digitization consisting of 7 ADCs. For every characterization, we take $Q=299$ samples from each ADC. All datasets are free and available online at \cite{google_drive_link}. 

			\subsection{$\textbf{D}_{\textbf{WSS}}$: DLWSS Dataset}
			The $\textbf{D}_{\textbf{WSS}}$ dataset is generated using the complex-valued normalized pseudo reconstructed signal, $\tilde{\textbf{X}}_{n}$ of size $N\times Q\times 2$.  For generating this dataset, real and complex parts of $\tilde{\textbf{X}}$ are separated and then normalized in the range $[0,1]$ as shown in Fig.~\ref{BD}. Since this dataset is used for spectrum sensing, the label, $\hat{\textbf{s}}$ of each frequency band will either be vacant (i.e. $\hat{s}_n = 0$) or busy (i.e. $\hat{s}_n = 1$).
			

			\subsection{$\textbf{D}_{\textbf{NMC}}$: NDLMC Dataset}
			The $\textbf{D}_{\textbf{NMC}}$ dataset uses $\hat{\textbf{X}}_{bb}$ which is generated by passing the reconstructed wideband signal via DDC.
			It is different compared to \cite{ref4,ref3} where single frequency band is considered compared to 14 bands simultaneously in our dataset depicting real wideband spectrum based ASCW. 
			Thus, the label is $\hat{\textbf{k}}$ of size $N\times 1$ where $n^{th}$ entry is $\hat{k}_n \in [0,M]$. Here, $\hat{k}_n=0$ denotes that $n^{th}$ frequency band is vacant and $\hat{k}_n\in[1,M]$ denotes that the $n^{th}$ frequency band is occupied with any one of the $M$ modulations schemes $\in [1,M]$.
			 For performance analysis in various scenarios, $\textbf{D}_{\textbf{NMC}}$ is further divided into two sections: 1) $\textbf{D}_{\textbf{NMC\_IQ}}$: Time domain IQ samples of pre-processed $\hat{\textbf{X}}_{bb}$, and 2) $\textbf{D}_{\textbf{NMC\_AP}}$: Time domain amplitude-phase vectors (i.e. polar representation of IQ samples) of pre-processed $\hat{\textbf{X}}_{bb}$. Each dataset has a shape of $N\times L\times 2$ where  $L = 256$ is the number of modulated symbols. In $\textbf{D}_{\textbf{NMC\_IQ}}$, the two vectors of third dimension denote in-phase and quadrature phase components of pre-processed $\hat{\textbf{X}}_{bb}$. We normalize this dataset in the range [0,1] before passing it to the deep learning models. Similarly, in $\textbf{D}_{\textbf{NMC\_AP}}$,  amplitude and phase form two vectors of third dimension. The amplitude is normalized via $l_2$ norm and phase (in radians) is normalized in the range [-1, 1] as in \cite{ref3}.
			
			Furthermore, in each case, we consider three types of wireless channels: 1) AWGN ($\textbf{D}_{\textbf{NMC\_IQ1}}$ and $\textbf{D}_{\textbf{NMC\_AP1}}$), 2) AWGN channel and Rayleigh fading with a Doppler shift of 10~Hz ($\textbf{D}_{\textbf{NMC\_IQ2}}$ and $\textbf{D}_{\textbf{NMC\_AP2}}$), and 3) AWGN channel and Rician fading with a Doppler shift of 10~Hz ($\textbf{D}_{\textbf{NMC\_IQ3}}$ and $\textbf{D}_{\textbf{NMC\_AP3}}$).

				\subsection{$\textbf{D}_{\textbf{WMC}}$: WDLMC Dataset}
			The $\textbf{D}_{\textbf{WMC}}$ dataset is generated directly from the complex valued reconstructed wideband signal, $\hat{\textbf{X}}$ of size $N\times Q$. This dataset is also separated into real and imaginary parts and hence, it is of size $N\times Q\times 2$. Since the dataset classifies the modulation schemes of all $N$ frequency bands, the dataset $\textbf{D}_{\textbf{WMC}}$ has labels, $\hat{\textbf{k}}\in[0,7]$ as in $\textbf{D}_{\textbf{NMC}}$. Likewise, we consider three types of wireless channels: 1) AWGN ($\textbf{D}_{\textbf{WMC1}}$), 2) AWGN channel and Rayleigh fading with a Doppler shift of 10~Hz ($\textbf{D}_{\textbf{WMC2}}$), and 3) AWGN channel and Rician fading with a Doppler shift of 10~Hz ($\textbf{D}_{\textbf{WMC3}}$).

			\subsection{Training Parameters and Tools} The neural networks are implemented using  Keras \cite{keras} (with Tensorflow backend \cite{ref31}) on Nvidia Cuda \cite{ref32} enabled Quadro P4000 GPU. The weights of the models are initialized using default Keras initializers. We use an Adam optimizer whose parameters are set as  $\beta_1 = 0.9$ and $\beta_2 = 0.999$ \cite{ref33}.

			\section{Performance Comparison}


		In this section, we present results to compare the performance of the proposed architectures with the state-of-the-art works in the literature. From a wireless communication perspective, we consider two parameters: 1) Spectrum sensing accuracy which in turn guarantees accurate spectrum reconstruction from SNS samples, and 2) Modulation classification accuracy for a wide range of SNRs and wireless channels.
		
		
		\subsection{Spectrum Sensing Performance Analysis}
\begin{figure}[!h]
	\vspace{-0.2cm}
			\centering
			{\includegraphics[scale=0.5]{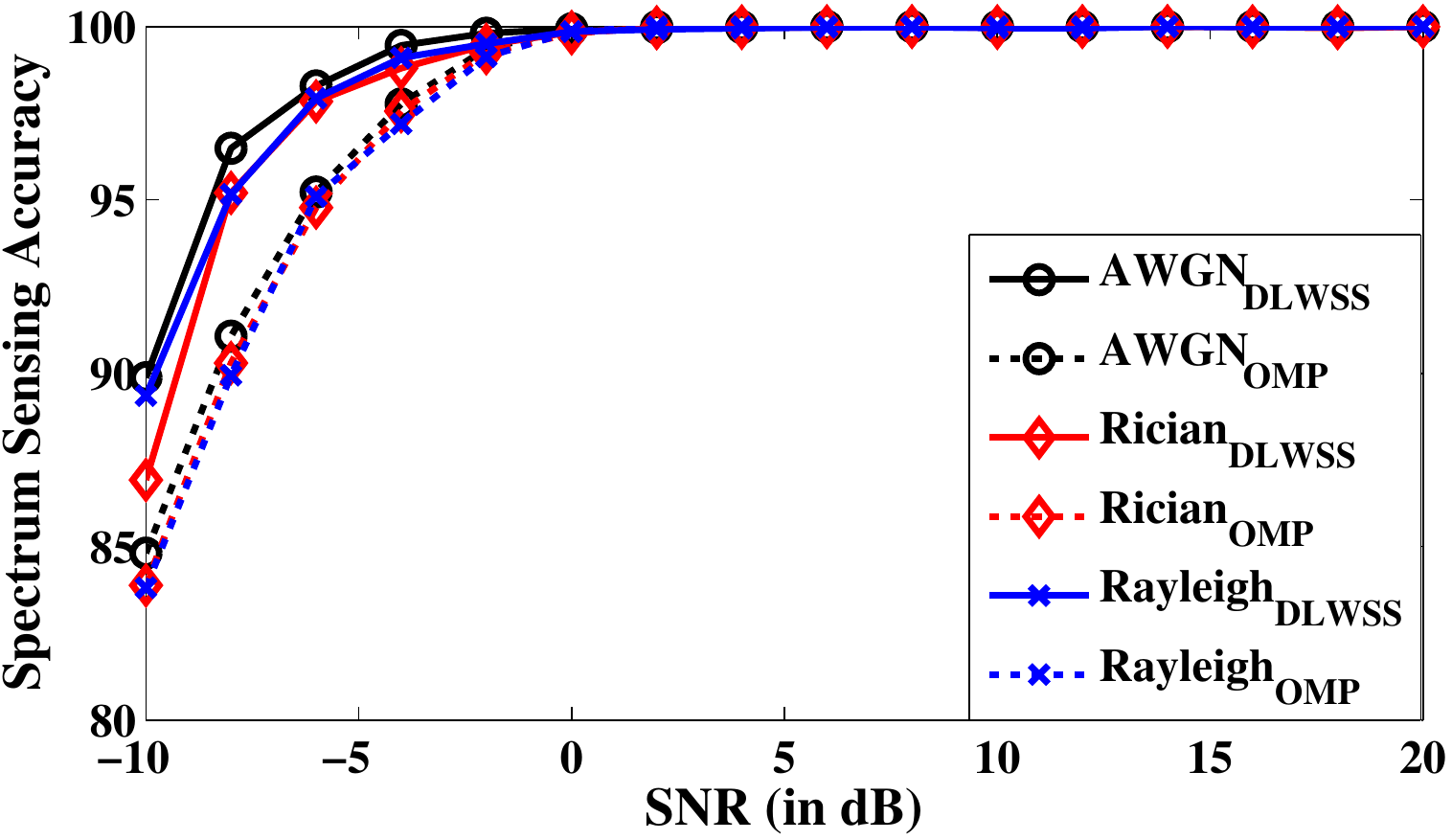}}
		\vspace{-0.2cm}
			\caption{Spectrum sensing accuracy of the proposed DLWSS and existing OMP method for various channel models.}
			\label{Pr_detect}
				\vspace{-0.2cm}
		\end{figure}

	In Fig.~\ref{Pr_detect}, we compare the spectrum sensing accuracy of the proposed DLWSS architecture with conventional OMP based approach \cite{omp} which unlike DLWSS, requires the prior knowledge of the number of busy frequency bands. The analysis is performed for three different wireless channels. As expected, DLWSS outperforms OMP at low SNRs. Furthermore, at a low SNR of $-10$~dB, the accuracy is {5.02}\%, {5.53}\% and {3.04}\% higher than OMP for AWGN, Rayleigh  and Rician fading channels, respectively. Thus, switching from OMP to DLWSS based spectrum reconstruction not only improves the performance but also allows a reconfigurable architecture via a single unified pipeline due to similarity between DLWSS, NDLMC and WDLMC building blocks.

		\begin{figure*}[!b]
			\centering
				\vspace{-0.2cm}
			\subfloat[]{\includegraphics[scale=0.5]{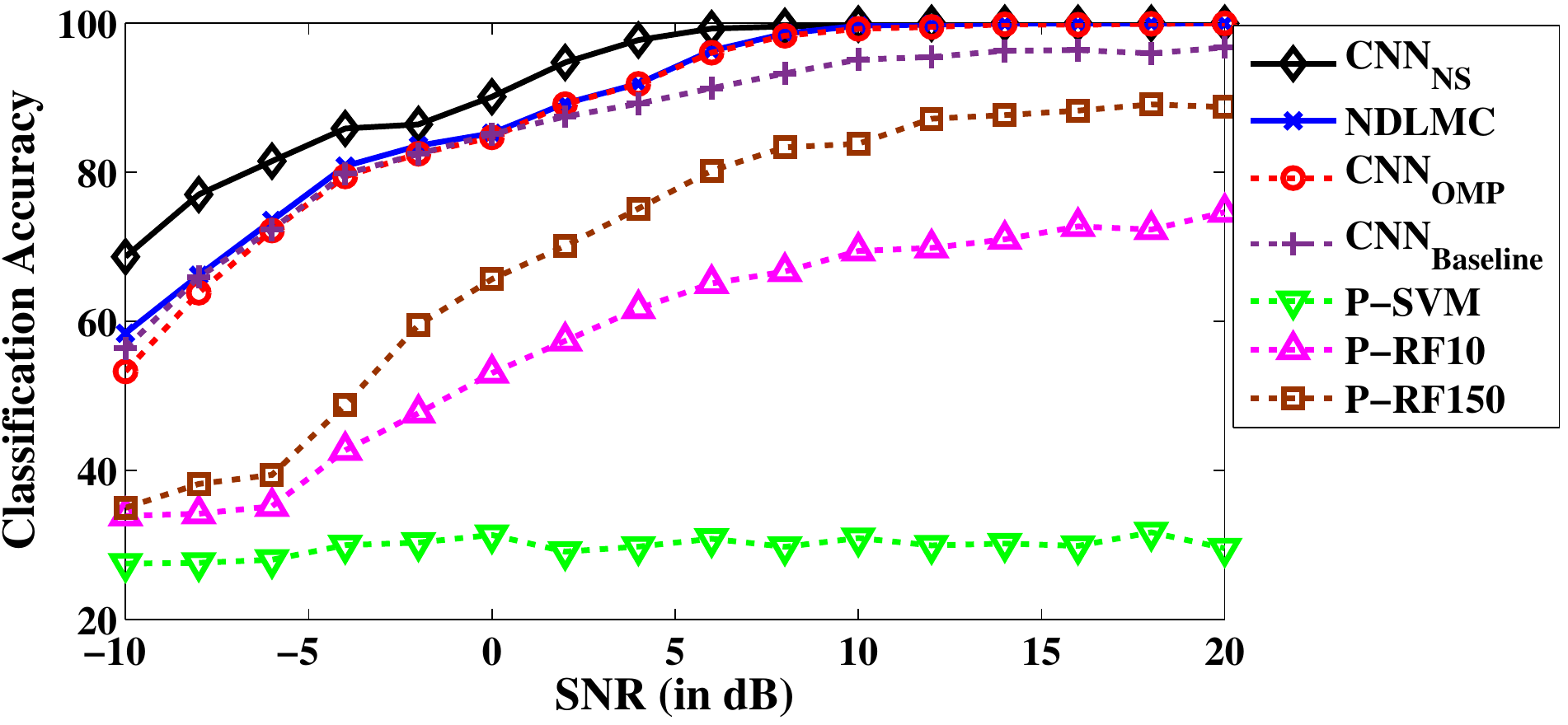}}
			\hspace{0.15cm}
			\subfloat[]{\includegraphics[scale=0.5]{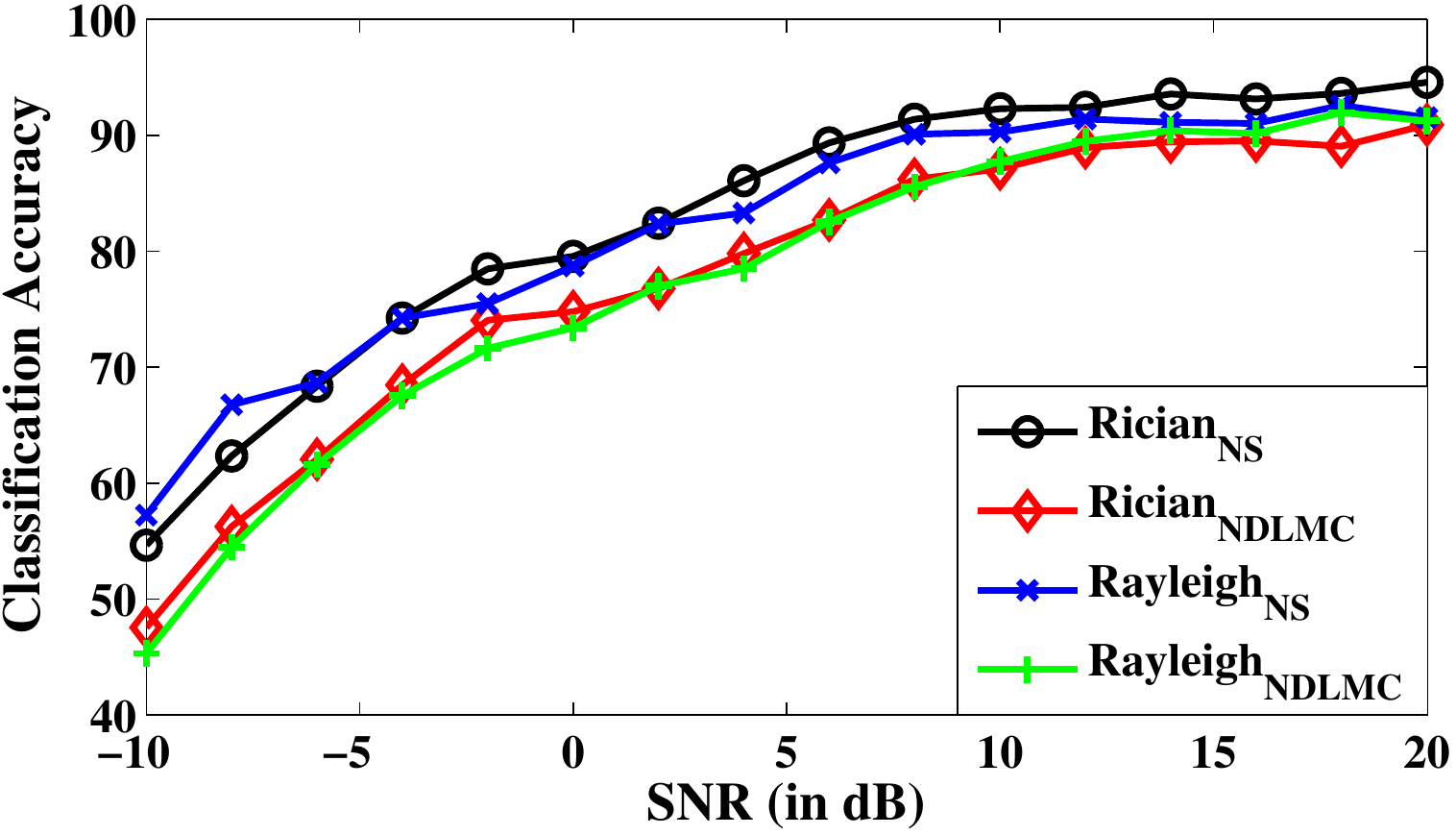}}
				\vspace{-0.2cm}
			\caption{Modulation classification accuracy of (a) NDLMC and other classification methods on the proposed ASCW for the dataset ~$\textbf{D}_{\textbf{NMC\_IQ1}}$ (i.e. IQ samples of AWGN channel)
			(b) NDLMC and NS based NDLMC on Rayleigh (i.e. dataset~$\textbf{D}_{\textbf{NMC\_IQ2}}$) and Rician (i.e. dataset~$\textbf{D}_{\textbf{NMC\_IQ3}}$) channel models.}

			\label{class_result}	
			\vspace{-0.2cm}
		\end{figure*}



		\subsection{Modulation Classification Accuracy Comparison}
\begin{figure*}[!htbp]
			\centering
				\vspace{-0.2cm}
			\subfloat[]{\includegraphics[scale=0.51]{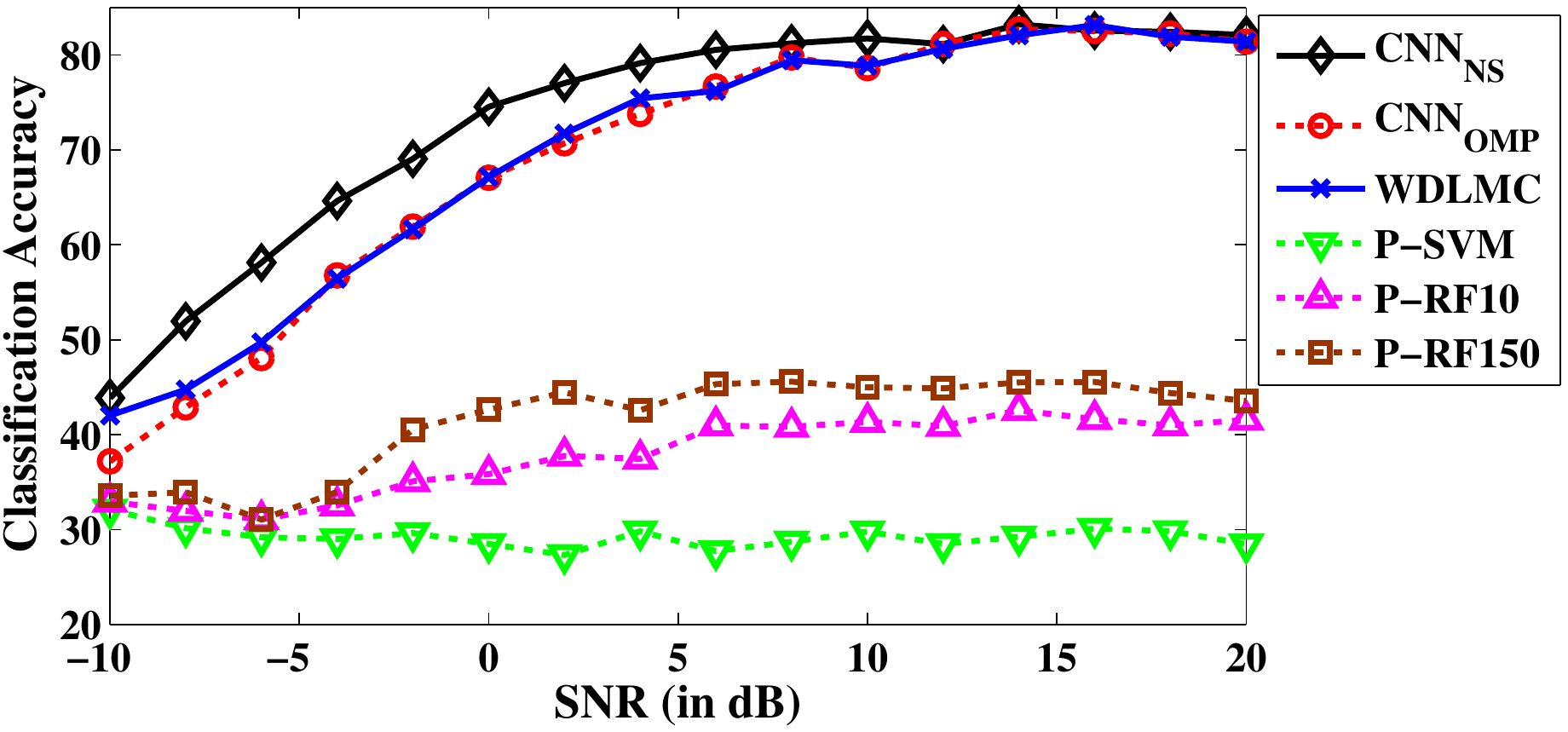}}
			\hspace{0.15cm}
			\subfloat[]{\includegraphics[scale=0.51]{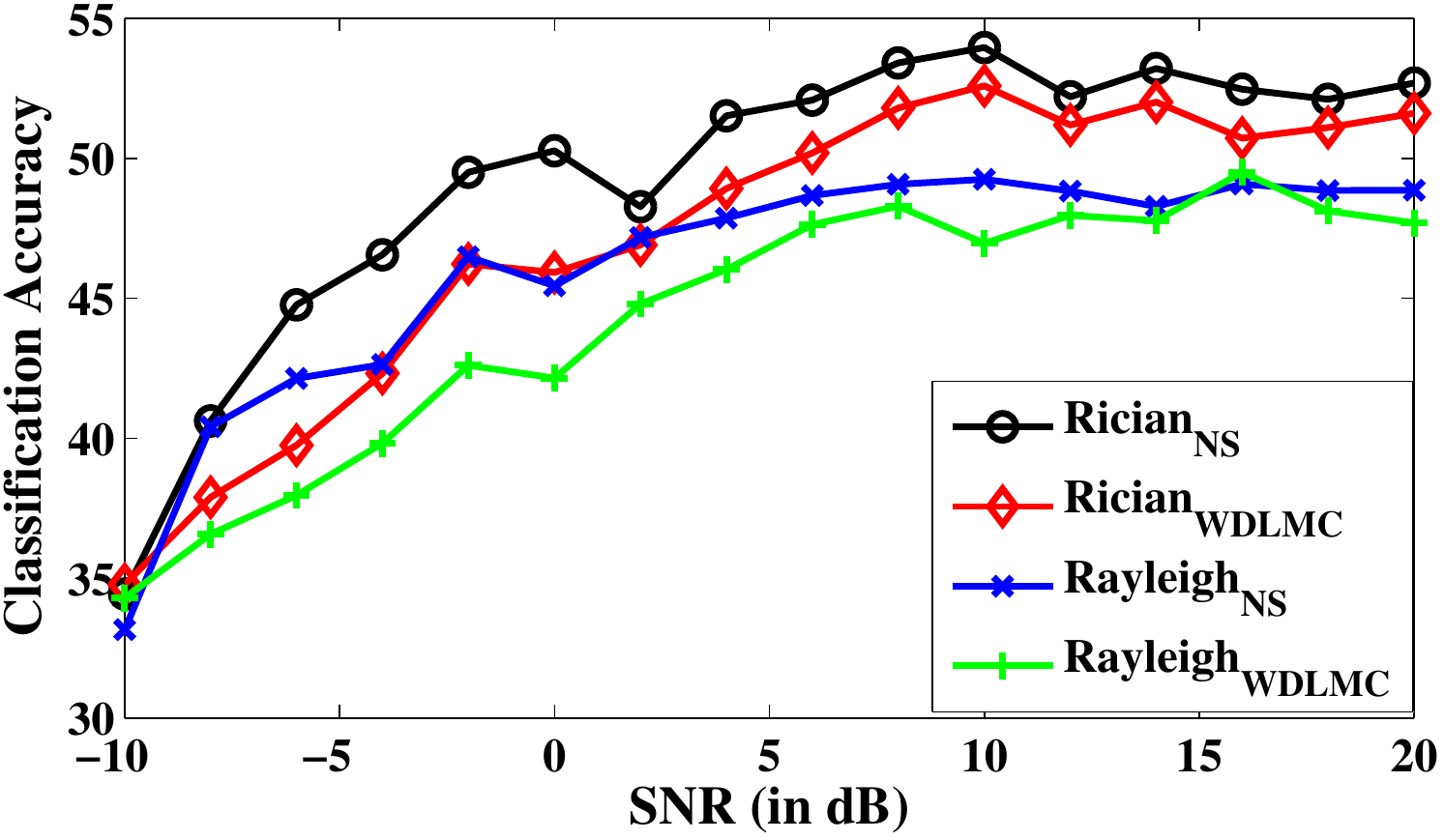}}
				\vspace{-0.2cm}
			\caption{Modulation classification accuracy of  (a) WDLMC and other classification methods on the proposed ASCW for the dataset ~$\textbf{D}_{\textbf{WMC1}}$ (i.e. AWGN channel) (b) WDLMC and NS based WDLMC on Rayleigh (i.e. the dataset~$\textbf{D}_{\textbf{WMC2}}$) and Rician (i.e. the dataset~$\textbf{D}_{\textbf{WMC3}}$) channel models.}

			\label{class_result_WDLMC}	
			\vspace{-0.2cm}
		\end{figure*}

		For modulation classification, we compare the  performance of the proposed NDLMC and WDLMC architectures for various IQ datasets discussed in Section VI. Please refer to Appendix for results on AP datasets.
		We consider the comparison with various approaches discussed in Section II such as SVM classifier with linear function kernel (P-SVM) and random forest with 10 and 150 trees in \cite{ref3,ref4} (P-RF10/150). Since these approaches demand Nyquist-sampled signal, we used the proposed DLWSS based DLDR for reconstruction. Furthermore, these approaches cannot characterize the multiband signal directly and hence; we sequentially pass the detected busy bands using the tunable DDC compared to simultaneous classification in the proposed approach. Thus, only busy bands were considered while calculating modulation classification accuracy. 
		In addition, we also considered the SVM classifier used in \cite{cssp}, but its performance is similar to P-RF150, and hence, it is not included in the plots. 
The	$CNN_{NS}$ approach in Fig.~\ref{class_result} and Fig.~\ref{class_result_WDLMC} are the proposed NDLMC and WDLMC applied directly on the baseband converted Nyquist samples (NS) and wideband NS, respectively. Also, $CNN_{NS}$ requires the prior knowledge of busy bands \cite{ref4}.
In contrast, $CNN_{OMP}$ represents NDLMC (for Fig.~\ref{class_result}) and WDLMC (for Fig.~\ref{class_result_WDLMC}), applied on the signal reconstructed via OMP instead of DLWSS. Note that the NS approach needs additional signal processing operation before classification.

		\subsubsection{$\textbf{D}_{\textbf{NMC\_IQ}}$}
		The modulation classification accuracy of baseband converted wideband signal for AWGN, Rayleigh and Rician channels for dataset  $\textbf{D}_{\textbf{NMC\_IQ}}$ are shown in Fig.~\ref{class_result} (a) and (b). 
		
		For AWGN channel with dataset $\textbf{D}_{\textbf{NMC\_IQ1}}$, as shown in Fig.~\ref{class_result}~(a), average accuracy is 88.95\% and 96.42\% for SNR range of -10~dB to 20~dB and 0~dB to 20~dB, respectively.  At high SNR, accuracy of the proposed architectures is same as NS based approaches while this is not true for $CNN_{Baseline}$ approach. As discussed before, $CNN_{Baseline}$ also demands additional complex signal processing between reconstruction and classification stages.

		Next, we consider the challenging Rayleigh and Rician wireless channels and corresponding classification accuracy for datasets $\textbf{D}_{\textbf{NMC\_IQ2}}$ and $\textbf{D}_{\textbf{NMC\_IQ3}}$, respectively, is shown in Fig.~\ref{class_result}~(b). Here ${Rayleigh}_{NS}$ and ${Rician}_{NS}$ use the CNN classifier on the baseband converted received wideband signal. Overall, average accuracy for Rayleigh and Rician fading channels is  77.40\% and 77.72\%, respectively along with closed match between SNS and NS approaches.
		
 	\begin{figure}[t]	
					\centering
					\subfloat[]{\includegraphics[scale=0.3750]{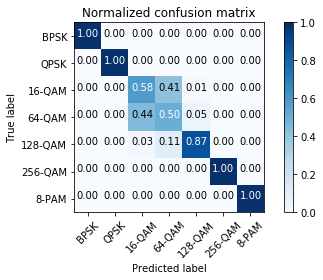}}
					\hspace{0.125cm}
					\subfloat[]{\includegraphics[scale=0.3750]{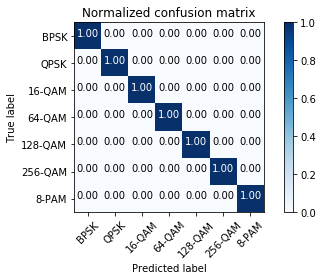}}\\

					\vspace{-0.2cm}	\subfloat[]{\includegraphics[scale=0.375]{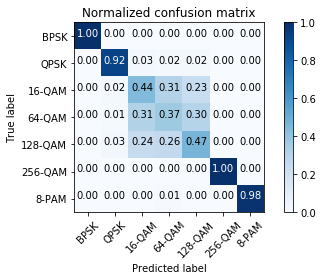}}
					\hspace{0.125cm}
					\subfloat[]{\includegraphics[scale=0.3750]{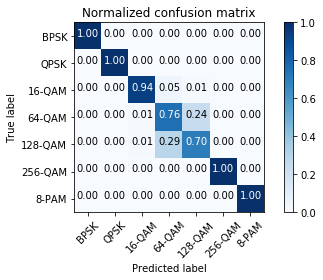}}\\
					\vspace{-0.2cm}	\subfloat[]{\includegraphics[scale=0.3750]{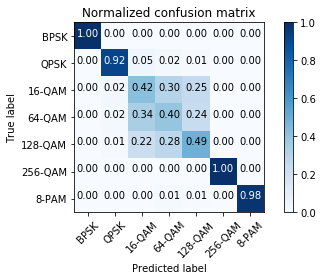}}
					\hspace{0.125cm}
					\subfloat[]{\includegraphics[scale=0.3750]{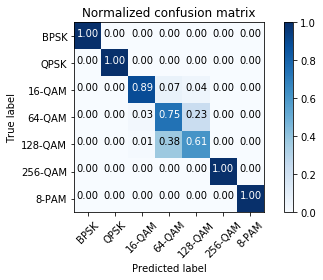}}

					\vspace{-0.2cm}
					\caption{Confusion plots of NDLMC on the proposed ASCW for (a) AWGN channel i.e. $\textbf{D}_{\textbf{NMC\_IQ1}}$ at SNR = 0~dB (b)AWGN channel i.e. $\textbf{D}_{\textbf{NMC\_IQ1}}$ at SNR = 18~dB (c) Rayleigh channel i.e. $\textbf{D}_{\textbf{NMC\_IQ2}}$ at SNR = 0~dB (d) Rayleigh channel i.e. $\textbf{D}_{\textbf{NMC\_IQ2}}$ at SNR = 18~dB (e) Rician channel i.e. $\textbf{D}_{\textbf{NMC\_IQ3}}$ at SNR = 0~dB (f) Rician channel i.e. $\textbf{D}_{\textbf{NMC\_IQ3}}$ at SNR = 18~dB. }
					\label{conf_plots_NDLMC}
				\end{figure}

		To understand the classifier performance and inter-class discrepancies better, we analyse the confusion plots of the proposed NDLMC at an SNR of 0~dB and 18~dB for all three channel models in Fig.~\ref{conf_plots_NDLMC}. For all the channel models, at an SNR of 18~dB, we can clearly see a sharp diagonal with almost perfect classification except for 16-QAM and 64-QAM. As the SNR reduces, the sharpness of the diagonal further reduces in the 16/64/128 QAM region. Since similar observation is also valid for NS based classifier, we observed that classification of QAM schemes at low SNR is challenging and there is a scope for improvement. Nonetheless, the proposed solution offers better performance than existing state-of-the-art approaches and directions for future work.
		

		\subsubsection{WDLMC}
		Next, we consider WDLMC scenario for the same three channels and corresponding results are shown in  Fig.~\ref{class_result_WDLMC}~(a) and (b) for dataset $\textbf{D}_{\textbf{WMC}}$.
		Since the classification of direct wideband signal has not been done yet in the literature, there is no baseline architecture for comparison. 
	 	For AWGN channel with dataset $\textbf{D}_{\textbf{WMC1}}$, as shown in Fig.~\ref{class_result_WDLMC}~(a), average accuracy is 70\% and 78\% for SNR range of -10~dB to 20~dB and 0~dB to 20~dB, respectively. At high SNR, accuracy of the proposed architectures is same as NS based approaches.

        Classification performance of dataset $\textbf{D}_{\textbf{WMC2}}$ and $\textbf{D}_{\textbf{WMC3}}$ is shown in Fig.~\ref{class_result}~(d). Here ${Rayleigh}_{NS}$ and ${Rician}_{NS}$ use the CNN classifier directly on the received wideband signal, $x(t)$.
		Although the performance of the proposed SNS based WDLMC approaches to that of Nyquist sampling based DLMC (i.e. ${Rayleigh}_{NS}$ and ${Rician}_{NS}$) but as expected the performance of both NS and SNS based modulation classifier is lower than baseband converted signal shown in Fig.~\ref{class_result}~(b). This is the small penalty incurred to reduce the complexity of additional signal processing between reconstruction and classification stages along with making the  architecture reconfigurable. 
	
	
\vspace{-0.3cm}

			\section{Conclusions and Future Directions}

            In this paper, we presented a novel deep-learning (DL) based framework for automatic spectrum characterization of wideband spectrum (ASCW). We discussed the need for ASCW in next-generation networks, sub-Nyquist sampling (SNS) along with the demand of reconfigurable architecture and presented a single unified pipeline based DL architecture for SNS receivers. The proposed architectures are based on the in-depth study of DL frameworks along with detailed experimental analysis over a wide range of wireless channels, signal-to-noise ratio (SNR) and modulation schemes. We not only presented a state-of-the-art architecture for conventional narrowband ASCW approach but also explored the characterization of direct wideband spectrum for the first time in the literature. 
            
            Our results highlighted that the performance of the wideband approach is lower than narrowband approach, and more research works on DL framework and SNS approach is needed. Furthermore, the characterization of direction-of-arrival, communication standard and jammer are also important areas to explore. Another future direction involves intelligent ASCW where additional intelligence is needed to identify the part of the spectrum to be digitized. This is important as expected operating frequency range in 5G and 6G networks is up to 60 GHz and 100 THz, respectively, and direct digitization of such wideband spectrum is challenging and inefficient. From an architecture perspective, mapping the proposed architecture to hardware, integration with SNS platform and meeting the desired latency constraints of next-generation networks are important and open research challenges.

			\appendices

  \section{LSTM Architecture for NDLMC}
  \label{lstm_ndlmc}
 

			\begin{figure}[!b]
						\centering
						\includegraphics[scale=0.55]{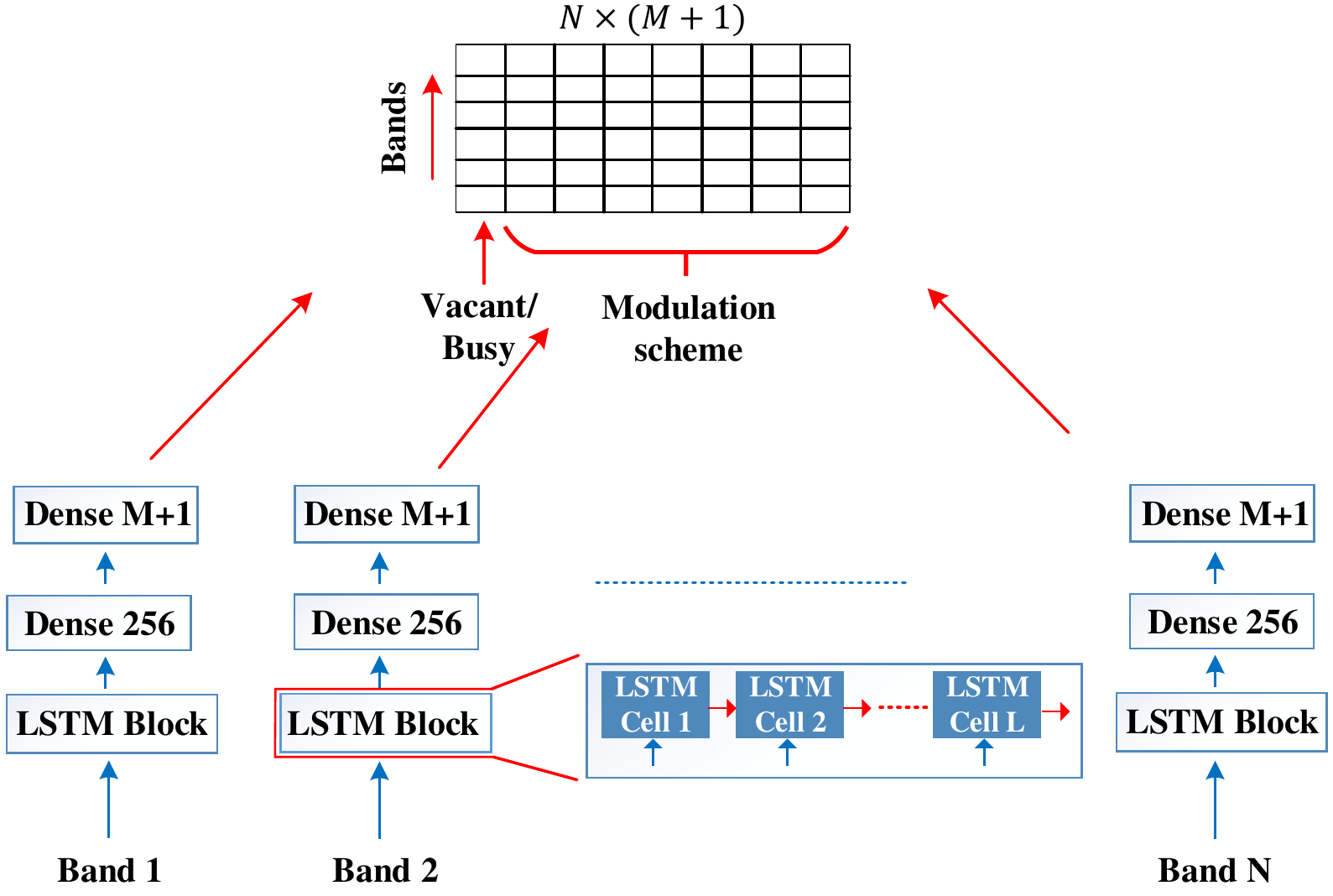}
						\vspace{-0.2cm}
						\caption{LSTM architecture for NDLMC.}
						
						\label{LSTM_architecture}
						\vspace{-0.2cm}
					\end{figure}
					
		Various studies in the literature have shown that the LSTM based architecture offers slightly better performance for datasets with AP samples \cite{ref3}. We have also explored LSTM architecture for NDLMC and WDLMC tasks with SNS based digitization. The proposed novel architecture, shown  in  Fig.~\ref{LSTM_architecture}, consists of  $N$ parallel (one for each frequency band) neural network based  prediction  modules and these modules share the learned parameters. Note that the \textit{LSTM block} in Fig.~\ref{LSTM_architecture} comprises of $L$ LSTM cells \cite{lstm}. Thus, the proposed architecture can directly process a signal of dimension $N\times L\times 2$. With reference to NS based LSTM classifier in \cite{ref3}, the proposed architecture can process the multi-band signal simultaneously without increasing the weight-complexity. 
		
\begin{figure}[!h]
				\centering
				\vspace{-0.2cm}
				\includegraphics[scale=0.51]{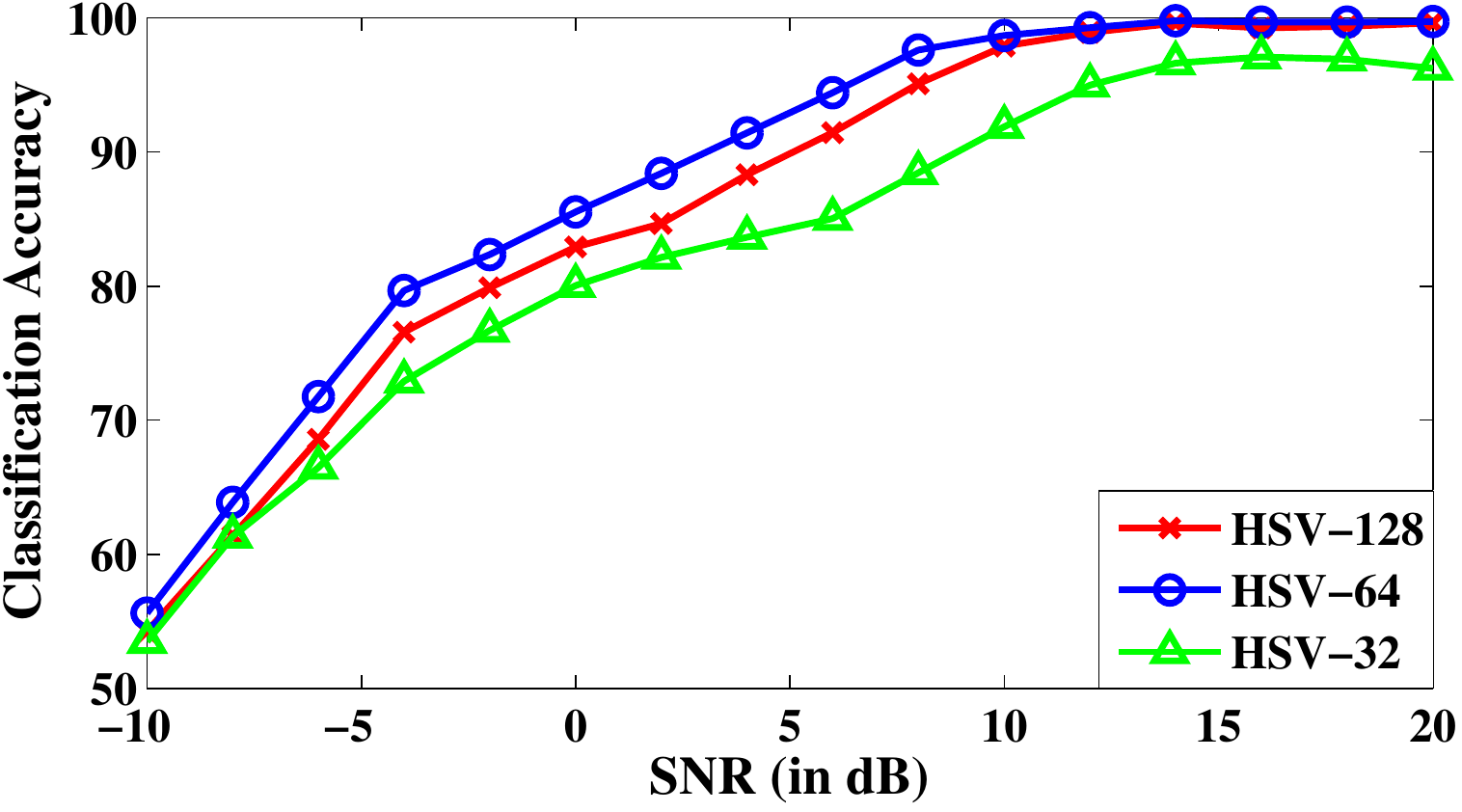}\vspace{-0.2cm}
				\caption{Ablation study of LSTM based NDLMC on AP samples of $\hat{\textbf{X}}_{bb}$.}
				\label{lstm_hsv}
			\end{figure}

		For this architecture, we perform the ablation study for selecting the appropriate value of hidden state vector (HSV) hyper-parameter \cite{ref3}. As shown in Fig.~\ref{lstm_hsv}, we choose HSV of 64 as it offers better performance than HSV = 32 and lower computational time than HSV = 128.
		
%
					
\begin{figure}[!h]
			\subfloat[]{\includegraphics[scale=0.45]{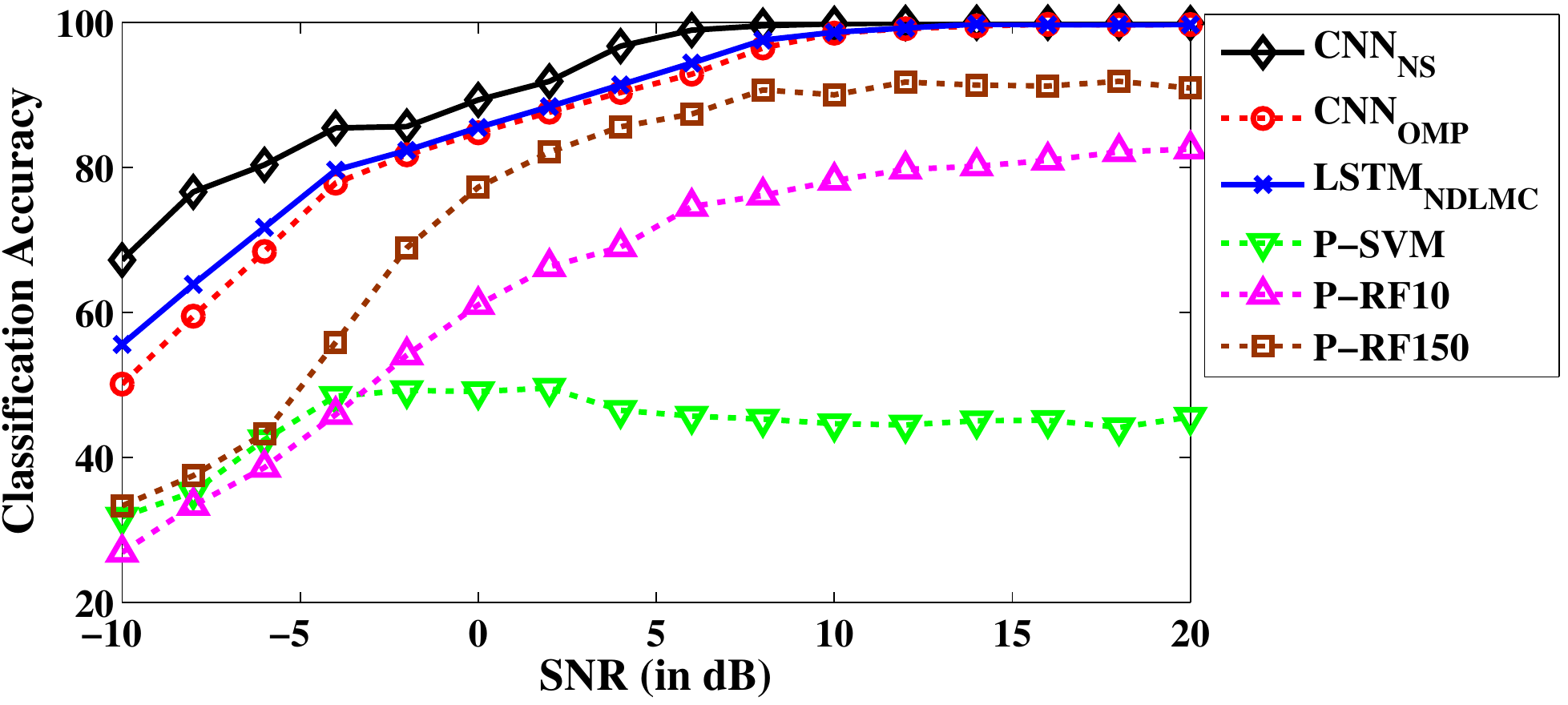}}\\
			\subfloat[]{\includegraphics[scale=0.45]{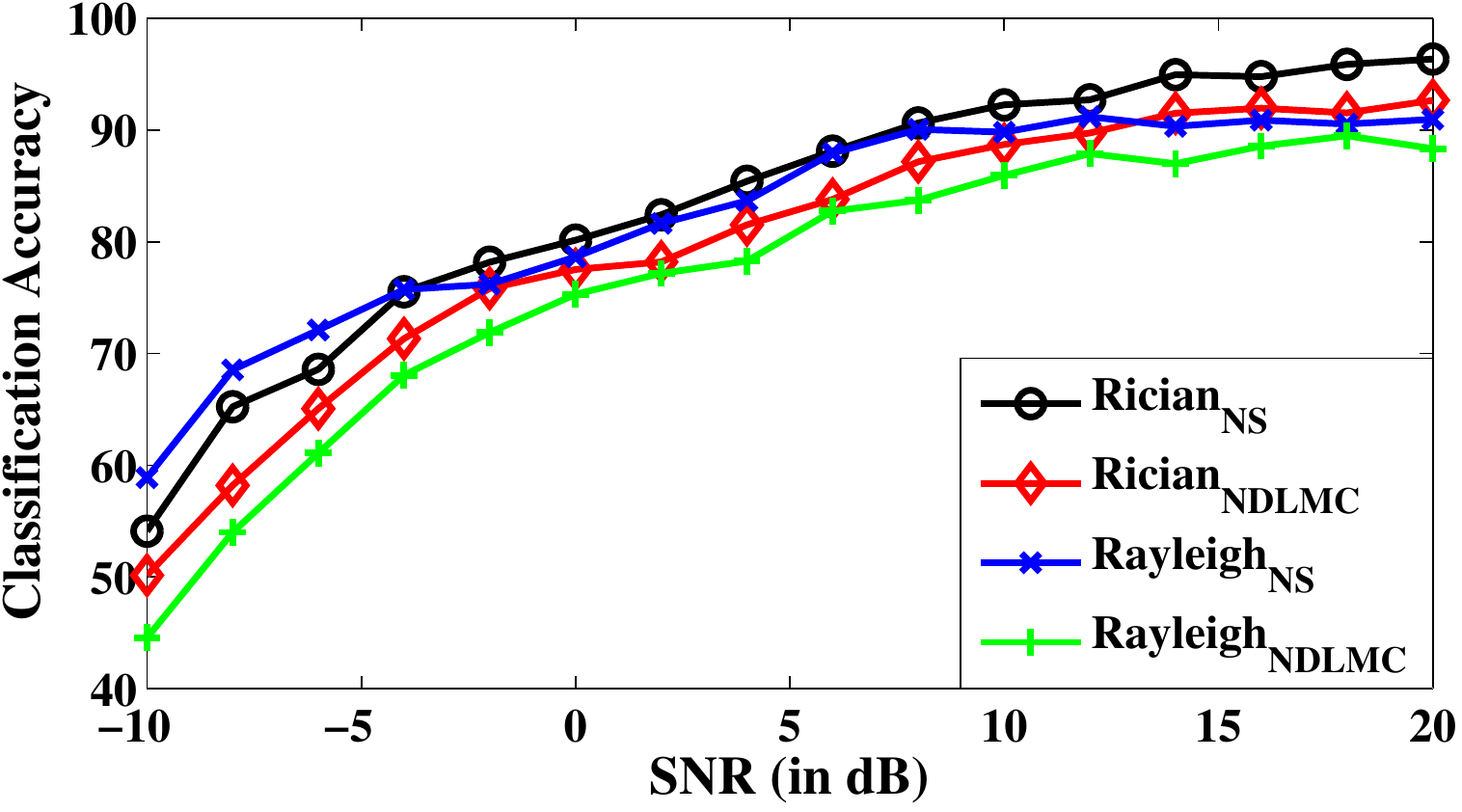}}
			\caption{Classification accuracy of LSTM based NDLMC on AP samples of (a)~AWGN channel i.e. $\textbf{D}_{\textbf{NMC\_AP1}}$ (b)~Rayleigh channel i.e. $\textbf{D}_{\textbf{NMC\_AP2}}$ and Rician channel $\textbf{D}_{\textbf{NMC\_AP3}}$.}
			\label{lstm_result}	
			\vspace{-0.1cm}
		\end{figure}

		
				The modulation classification accuracy of the proposed NDLMC on the LSTM classifier for AWGN, Rayleigh and Rician  fading channel models are shown in Fig.~\ref{lstm_result} (a) and (b).
					For AWGN channel with dataset $\textbf{D}_{\textbf{NMC\_AP1}}$, as shown in Fig.~\ref{lstm_result}~(a), the average accuracy is 87.96\% and 95.83\% for SNR range of -10~dB to 20~dB and 0~dB to 20~dB, respectively and similar to NDLMC on the CNN classifier, at a SNR of 10~dB, its classification accuracy becomes 100\%.
		        Fig.~\ref{lstm_result}~(b) shows the classification accuracy for Rayleigh and Rician fading channels i.e. for dataset $\textbf{D}_{\textbf{NMC\_AP2}}$ and $\textbf{D}_{\textbf{NMC\_AP3}}$, respectively. 
				The average accuracy for Rayleigh fading channel conditions is  76.4\% whereas it is 79.66\% for Rician fading channel. Please note that here ${Rayleigh}_{NS}$ and ${Rician}_{NS}$ use the LSTM classifier of architecture shown in Fig.~\ref{LSTM_architecture} on the baseband convertered $\hat{\textbf{X}}_{bb}$.
		
			 Table~\ref{lstm_comp} shows the classification performance comparison of NDLMC when CNN classifier and LSTM classifier are used for DLMC task. It is observed that the classification accuracy of NDLMC with CNN and LSTM classifier is almost same for AWGN and Rayleigh fading channel whereas LSTM classifier performs better than CNN classifier for Rician fading channel.

	 		\renewcommand{\thetable}{VII}	
			 	\begin{table}[!h]

					\begin{center}
						\caption{Average Classification performance Comparison of Signal recovered using proposed method. }
						\label{lstm_comp}
						\centering
						\renewcommand{\arraystretch}{1}
							\begin{tabular}{|c|c|c|c|}
								\hline
								\textbf{Method} & \multicolumn{1}{c|}{$\textbf{AWGN}$} & \multicolumn{1}{c|}{$\textbf{Rayleigh}$} & \multicolumn{1}{c|}{$\textbf{Rician}$}\\
								\cline{1-4}
								
															\textbf{NDLMC-LSTM}(${D}_{{NMC\_AP1}}$) &88&76.4 & 79.66
															\\	\hline
								\textbf{NDLMC-CNN}(${D}_{{NMC\_AP1}}$) &88.10&76.2 & 76.8
															\\	\hline
															
							\end{tabular}
					\end{center}
				\end{table}
			 

		  \section{LSTM Architecture for WDLMC}
\begin{figure}[!h]
				\centering
				\vspace{-0.2cm}
				\includegraphics[scale=0.48]{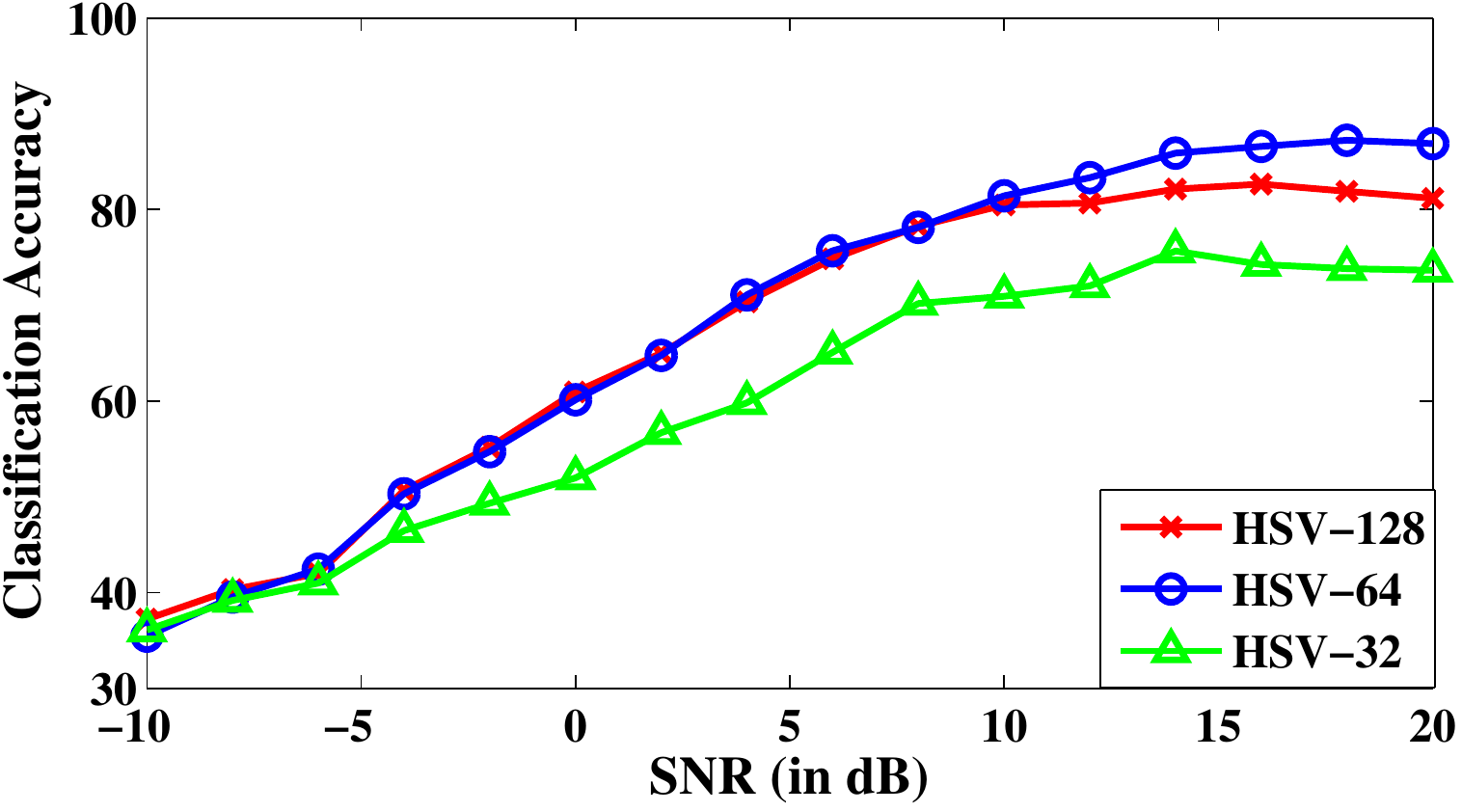}\vspace{-0.2cm}
				\caption{Ablation study of LSTM based WDLMC on $\hat{\textbf{X}}$ i.e. dataset $\textbf{D}_{\textbf{WMC1}}$. }
				\label{abl_raw_lstm}
			\end{figure}

		We use the LSTM architecture shown in Fig.~\ref{LSTM_architecture} for WDLMC. However, every LSTM block consists of Q LSTM cells. Thus, the architecture takes the direct wideband signal of dimension $N\times Q\times 2$ as input. We perform ablation study for the different size of HSV as shown in  Fig.~\ref{abl_raw_lstm}. It is found that similar to the NDLMC case, HSV=64 gives the better performance.

		  \subsection{Performance Analysis}
\begin{figure}[!h]
				\centering
				\vspace{-0.2cm}
				\includegraphics[scale=0.45]{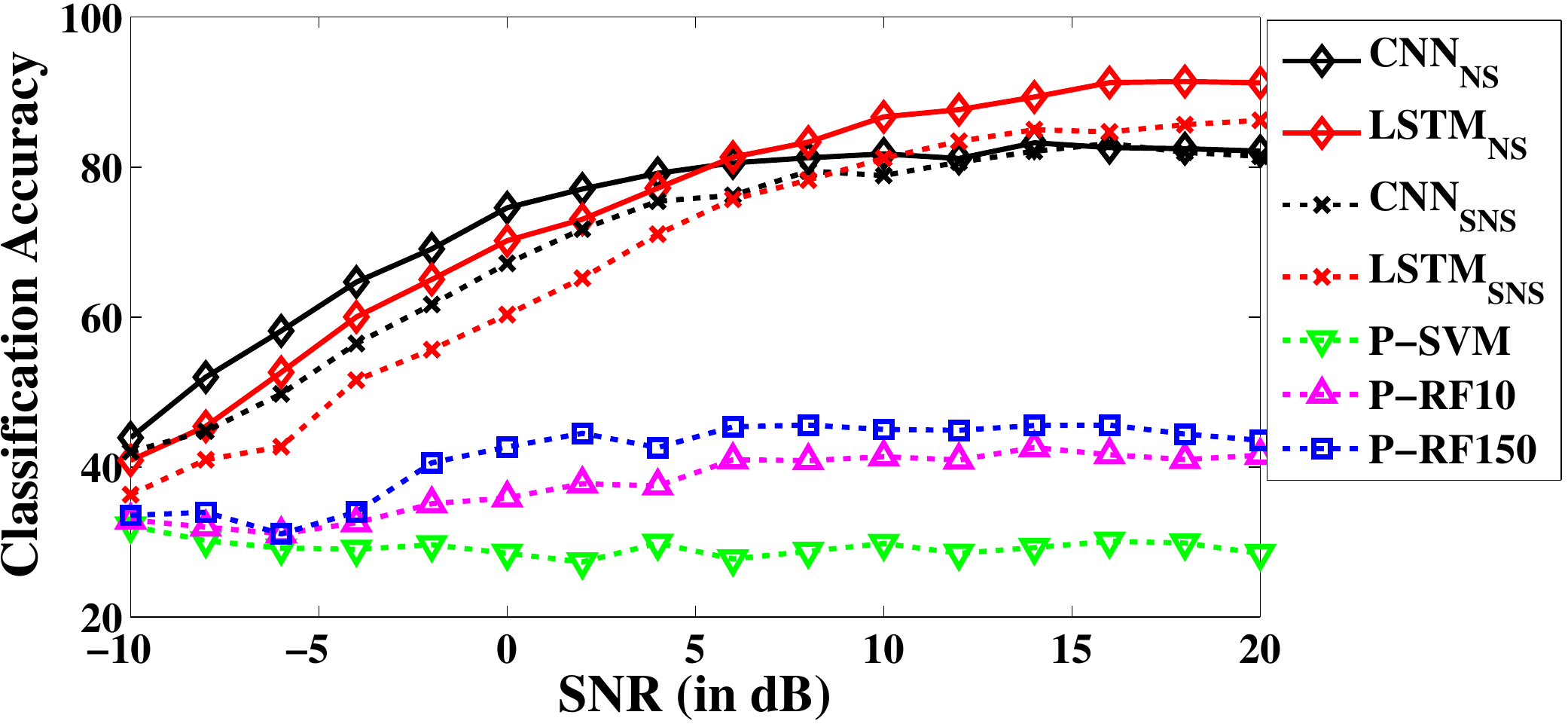}\vspace{-0.2cm}
				\caption{Classification Accuracy of CNN based WDLMC (i.e. $CNN_{NS}$ and $CNN_{SNS}$) and LSTM based WDLMC (i.e. $LSTM_{NS}$ and $LSTM_{SNS}$) on AWGN channel i.e. dataset $\textbf{D}_{\textbf{WMC1}}$.}
				\label{raw_cnn_lstm}
			\end{figure}

		  Next we show the classification performance comparison of LSTM with other algorithms on the reconstructed  wideband  signal, $\hat{\textbf{X}}$. Here, $LSTM_{SNS}$ and $CNN_{SNS}$ represent the performance of the LSTM and CNN, respectively, on the $\hat{\textbf{X}}$. For the AWGN  channel  with  dataset $\textbf{D}_{\textbf{WMC1}}$,  as  shown  in Fig.~\ref{raw_cnn_lstm} the average accuracy is 67.7\% for  SNR range of -10~dB to 20~dB.
		  Note that the performance of the proposed WDLMC-CNN model is better than the WDLMC-LSTM especially at low SNR. At very high SNR i.e above 12~dB, the performance of WDLMC-LSTM is found to be slightly better than WDLMC-CNN. However,the average accuracy of WDLMC-LSTM for  SNR range of -10~dB to 20~dB (i.e 67.7\%) is poorer than that of the WDLMC-CNN model (i.e 69.5\%).
		  
		  To summarize, LSTM based AMC does not offer significant improvement in performance compared to CNN based AMC in case of dataset with AP samples. Furthermore, LSTM performs poorly for IQ datasets. Thus, we conclude that CNN based classifier is the preferred approach for SNS based ASCW and offers reconfigurable architecture due to similarity with DLWSS block. 
		  
				\end{document}